\documentclass[aps,prx,twocolumn,superscriptaddress]{revtex4-2}
\usepackage[a4paper,margin=1.25cm]{geometry}
\usepackage{times}
\usepackage{graphicx}
\usepackage[font={small}]{caption}
\usepackage{subcaption}
\usepackage{amsmath}
\usepackage{amssymb}
\usepackage{tabularx}
\usepackage{placeins}
\usepackage{multirow}
\usepackage{braket}
\usepackage{color}
\usepackage{hyperref}
\usepackage[normalem]{ulem}
\usepackage{breakcites}
\usepackage{adjustbox}
\usepackage{mathtools}
\usepackage{float}
\usepackage{adjustbox}
\usepackage{dcolumn}% Align table columns on the decimal point
\usepackage{bm}% bold math
\usepackage{comment}
\usepackage{titlesec}
\usepackage{booktabs}
\usepackage{siunitx}
\usepackage{comment}
\usepackage{placeins}
\usepackage{graphicx}
\usepackage{subcaption}

\usepackage{graphicx}
\usepackage{subcaption}
\usepackage{placeins}
\usepackage{comment}

\usepackage[dvipsnames]{xcolor}

\titleformat{\subsubsection}{\normalfont\centering}{\thesubsubsection.}{1em}{}

\newcommand{\xcheck}[1]{\textbf{\color{red}[XX: check]}}

\newcommand{\aqcom}[1]{}

\newcommand{\Ticctqz}{-58.07270087}
\newcommand{\Ticctfivez}{-58.08119845}
\newcommand{\Ticctcbs}{-58.09078(81)}

\newcommand{\Ticcqcbs}{-58.09241(80)}
             
\newcommand{\Tistep}{-58.09145(10)}
\newcommand{\Tivar}{-58.09297(19)}
\newcommand{\Tieot}{-58.090754(42)}

\newcommand{\Nicctqz}{-169.3357012}
\newcommand{\Nicctfivez}{-169.3582773}
\newcommand{\Nicctcbs}{-169.38885(63)}

\newcommand{\Niccqcbs}{-169.38830(64)}
             
\newcommand{\Nistepmoredet}{-169.38533(37)}
\newcommand{\Nieotmoredet}{-169.37924(19)}

\newcommand{\TiHcctqz}{-58.65043884}
\newcommand{\TiHcctfivez}{-58.65931705}
\newcommand{\TiHcctcbs}{-58.66949(71)}
               
\newcommand{\TiHccqcbs}{-58.67593(70)}
               
\newcommand{\TiHstep}{-58.67112(13)}
\newcommand{\TiHvar}{-58.67362(37)}
\newcommand{\TiHeot}{-58.670359(55)}
               
\newcommand{\NiHcctqz}{-169.9430187}
\newcommand{\NiHcctfivez}{-169.9666837}
\newcommand{\NiHcctcbs}{-169.99856(56)}
               
\newcommand{\NiHccqcbs}{-169.99985(56)}

\newcommand{\NiHstepmoredet}{-169.99354(35)}
\newcommand{\NiHeotmoredet}{-169.98598(16)}

\newcommand{\BETiHcctqz}{197.89}
\newcommand{\BETiHcctfivez}{198.89}
\newcommand{\BETiHcctcbs}{200.4(2.8)}

\newcommand{\BETiHccqcbs}{213.1(2.8)}

\newcommand{\BETiHvar}{205.53(1.1)}
\newcommand{\BETiHstep}{202.96(0.42)}
\newcommand{\BETiHeot}{202.79(0.18)}

\newcommand{\BENiHcctqz}{269.6}
\newcommand{\BENiHcctfivez}{272.5}
\newcommand{\BENiHcctcbs}{275.9(2.2)}

\newcommand{\BENiHccqcbs}{280.7(2.2)}

\newcommand{\BENiHstepmoredet}{271.9(1.3)}
\newcommand{\BENiHeotmoredet}{268.08(65)}

\newcommand{\tihcrc}{204.6(8.8)}
\newcommand{\nihcrc}{240(8)}

\begin{document}
%~~~~~~~~~~~~~~~~~~~~~~~~~~~~~~~

\title{Toward Predictive Hydride Bond Energetics with Neural-Network Wavefunctions}

\thanks{
  Notice: This manuscript has been authored by UT-Battelle, LLC, under contract DE-AC05-00OR22725 with the US Department of Energy (DOE). The US government retains and the publisher, by accepting the article for publication, acknowledges that the US government retains a nonexclusive, paid-up, irrevocable, worldwide license to publish or reproduce the published form of this manuscript, or allow others to do so, for US government purposes. DOE will provide public access to these results of federally sponsored research in accordance with the DOE Public Access Plan (\url{http://energy.gov/downloads/doe-public-access-plan}).
}
%~~~~~~~~~~~~~~~~~~~~~~~~~~~~~~~
\author{Aqsa Shaikh}
\affiliation{Department of Physics, North Carolina State University, Raleigh, North Carolina 27695-8202, USA}
\author{Lubos Mitas}
\affiliation{Department of Physics, North Carolina State University, Raleigh, North Carolina 27695-8202, USA}
\author{P. Ganesh} 
\affiliation{Center for Nanophase Materials Sciences, Oak Ridge National Laboratory, Oak Ridge, Tennessee 37831, USA}
\author{Jaron T. Krogel}
\affiliation{Materials Science and Technology Division, Oak Ridge National Laboratory, Oak Ridge, Tennessee 37831, USA}
%\email{lmitas@ncsu.edu}
%~~~~~~~~~~~~~~~~~~~~~~~~~~~~~~~
\begin{abstract}{\centering}

Accurate prediction of transition-metal hydride (TM-H) bond dissociation energies (BDEs) remains challenging because of strong electron correlation, relativistic effects, and nuclear quantum contributions. Recent advances in neural-network wavefunctions have enabled highly accurate electronic structure calculations, yet their performance for hydride bond energetics has not been systematically benchmarked. Here, we assess the performance of neural-network variational Monte Carlo (NN-VMC) based on the Psiformer ansatz for systems with increasing complexity (LiH, OH, TiH and NiH), and compare it against coupled-cluster singles, doubles, triples, and perturbative quadruples with complete basis set extrapolation (CCSDT(Q)/CBS) as well as available theoretical and experimental data. Throughout this study, both NN-VMC and \textit{ab initio} calculations employ a common correlation-consistent effective core potential (ccECP) Hamiltonian to enable tractable and consistent comparisons. To reduce finite-training errors, we introduce complementary zero-variance and infinite-step extrapolation schemes.
For LiH and OH systems, NN-VMC yields total energies that differ within sub-milli Hartree compared to CBS extrapolated \textit{ab initio} results and the BDE differences remain under 2$\sigma$. In case of the early transition-metal systems Ti and TiH, the variational NN-VMC energies at the end of training are already consistent with CCSD(T)/CBS, while post-training extrapolation systematically closes the gap towards the CCSDT(Q)/CBS results. In contrast, NiH provides a stringent test of wavefunction expressivity, where increasing the number of determinants in NN-wavefunction ansatz, by going from 16 determinants to 160, improves the recovered correlation energy at the end of training by 14 mHa. The comparison of NiH BDE further reveals that, while the existing theoretical predictions cluster into distinct high and low BDE groups, the broad scatter and large uncertainties in available experimental data prevents a definitive assessment of the most accurate theoretical approach for NiH. Overall, this work demonstrates that NN-VMC with ccECPs Hamiltonian provides a competitive framework for quantitative prediction of main-group and early transition-metal hydride energetics, while identifying late transition-metal hydrides as an important benchmark for future developments in neural-network wavefunctions and electronic structure theory.
\end{abstract}
%%%~~~~~~~~~~~~~~~~~~~~~~~~~~~~~~~~~~~~~~~~~~~~~~~~~~~
\maketitle
%%%~~~~~~~~~~~~~~~~~~~~~~~~~~~~~~~~~~~~~~~~~~~~~~~~~~~

%%%~~~~~~~~~~~~~~~~~~~~~~~~~~~~~~~~~~~~~~~~~~~~~~~~~~~
\section{Introduction}\label{intro}
%%%~~~~~~~~~~~~~~~~~~~~~~~~~~~~~~~~~~~~~~~~~~~~~~~~~~~
Computational tools based on machine learning (ML) approaches such as neural networks (NN) and related methodologies have opened new perspectives for  many-body electronic structure calculations ~\cite{Carleo2017Science, Pfau-2020-ferminet, Hermann-2020-paulinet, vonGlehn-2022-psiformer}. These approaches have produced remarkably accurate results for several types of atomic, molecular, condensed and model systems ~\cite{choo-2020-fermionicNQS, scherbela-2022-weightsharing-mol, li-2022-deepsolid, gerard-2022-goldstandard, li-2022-ccECP-ML, pfau-2024-excited,scherbela-2024-transferable,szabo-2024-excitedstatevmc, scherbela-2025-deeperwin}. In some of these cases accuracy proved to be competitive with the most accurate correlated wave function methods that exploit basis set expansions such as Configuration Interaction and large-scale Coupled Cluster. Similarly, these approaches were on par or improved upon well-known  stochastic approaches such as quantum Monte Carlo (QMC) ~\cite{cassella-2023-discovering, ren-jichen-2023-NN-vmc-dmc,kessler-2021-ANNQMC}. In fact, many aspects of these NN/ML based methods are shared with variational Monte Carlo (VMC) ~\cite{ceperley-1977-fermionvmc} that has been used for decades ~\cite{ceperley-1980-electrongas, umrigar-1988-optimized, foulkes-2001-qmcreview, needs-2020-CASINO}.

The new NN approaches employ multi-layer functional architectures that vastly expand the variational freedom for description of many-body quantum effects. One of the advances is based on multi-particle/all-particle orbitals that are subsequently used to build the antisymmetric forms such as generalized many-body Slater determinants~\cite{Pfau-2020-ferminet, Hermann-2020-paulinet, vonGlehn-2022-psiformer}.  This provides a qualitative step forward in expanding  the variational freedom at the orbital level.
The second important point is the vast increase in the number of variational parameters that are organized into multi-layer, effectively nonlinear functional forms that are adaptive along the training/optimization evolution. Since this creates a nonlinear and hierarchical description, the  estimation of expectations is practically possible only in a stochastic manner, in effect following VMC approaches with sampling corresponding to the electronic degrees of freedom.
Optimization of such flexible forms have been shown to be highly effective in a number of cases, as cited before, with calculations now feasible for systems within moderate size limits ($\sim$200 electrons at present).
However, systematic comparisons between NN-VMC and other high-level wavefunction benchmarksunder consistent basis set convergence, pseudopotential protocols, and controlled extrapolation, remain limited.

Several recent benchmark studies have shown that transition metal (TM) systems are one of the difficult testing grounds for a variety of widely used electronic structure methods. Aoto et al. \cite{aoto-2017-icMRCC} produced a 60-molecule TM diatomic benchmark set using a composite core-valence coupled-cluster with singles, doubles, perturbative triples at complete basis set limit (CCSD(T)(CV)/CBS) along with an internally contracted multireference correction, $\Delta_{\rm MR}$, from icMRCCSD(T). Their main conclusion is not simply that coupled cluster fails for transition metals, but rather that carefully extrapolated CCSD(T) along with relevant corrections, such as spin-orbit (SO) and relativistic treatments, can be reliable in many cases, while some hydrides remain problematic even after multireference corrections. In particular, TiH and NiH were included among the hydrides for which the calculated bond dissociation energy (BDE) values are systematically higher than experiment by several kcal/mol; for NiH the disagreement was especially large, about 12 kcal/mol.
This is important because the reported $\Delta_{\rm MR}$ corrections for TiH and NiH are sizeable ($\approx$ 3-3.7 kcal/mol) but do not remove the discrepancy, suggesting that, in their study, the disagreement is not easily explained by a simple missing multireference correction alone.
Shee and co-workers \cite{Shee-2019-AFQMC} later examined 44 first row TM diatomics with phaseless Auxiliary Field quantum MonteCarlo (ph-AFQMC) and compared directly to Density Functional theory (DFT), CCSD(T), and multi-reference (MR) CCSD(T) data. Their ph-AFQMC results showed an improved overall consistency with experiment, but also reinforced the special difficulty of certain hydrides, including NiH. For TiH the ph-AFQMC estimate lies closer to the selected experimental value than the MR-CCSD(T) result of Aoto et al., whereas NiH required their most complete ph-AFQMC/CBS treatment, where ph-AFQMC at triple- and quadruple-zeta (TZ and QZ) basis sets levels were used to extrapolate to CBS value, unlike for TiH where CBS correction was scaled using coupled-cluster at perturbative triples.
Notably, Moltved and Kepp identified a broader \textit{metal hydride problem} by studying neutral and cationic M--H bonds across the transition series \cite{moltved-2019-ccsdt}. Their study includes testing various DFT functionals as well as CCSD(T) with TZ basis set. They trace much of the DFT difficulty to self interaction error and diffuse hydride-like density, while also noting that single-reference post-HF methods such as CCSD(T) can behave unexpectedly poorly for these bonds. In their analysis, NiH is one of the pathological CCSD(T) cases, further supporting the view that NiH is not a routine single-reference benchmark. The aforementioned studies also emphasized on the need for a revised experimental data for NiH. Taken together, these studies motivate the present focus on TiH and NiH,  providing an early resolvable TM-H, TiH, and a more challenging case, NiH, with persistent discrepancies among experiment and various theoretical results.

In this contribution we first address the technicalities just mentioned such as basis set extrapolations, convergence protocols, etc., so that we can start with a clean slate for more salient aspects of the considered methods. 
We examine zero-variance \cite{fu-ji-chen-2024-var-ext} and infinite-training extrapolations \cite{scherbela-2025-inv-step} recently used to obtain asymptotic predictions for NN-VMC energies within a particular ansatz.  We develop a robust regression approach in order to control the outliers and to allow for a direct assessment of sensitivity to various choices in the fitting process further discussed in \ref{subsec: Extrapolation-detailed}.
We show that, for the studied systems, robust regressions toward zero-variance and infinite-training produce BDEs in  quantitative agreement, despite the distinct assumptions intrinsic to each approach.  
While the primary objective of this work is to assess the expressivity of neural-network wavefunctions for transition-metal systems, we begin with two comparatively simple \textit{sp}-block molecules, LiH and OH. These systems serve as useful reference cases, as the different electronic-structure approaches considered here yield results that are statistically indistinguishable within their respective numerical uncertainties. For LiH, we quantify the impact of pseudopotential-related errors present in both the NN-VMC and other high-level \textit{ab initio} calculations. We then extend the analysis to the early transition-metal hydride TiH, where the NN-VMC total energies of both Ti and TiH reach the CCSD(T)/CBS level by the end of training, while the extrapolated estimates approach the CCSDT(Q)/CBS benchmark. In contrast, NiH presents a substantially greater challenge. Despite the use of extrapolation techniques, the NN-VMC calculations, with the same parameters as for the other systems, remain approximately 2$\%$ short of the correlation energy recovered by CCSDT(Q)/CBS. To better understand this discrepancy, we investigate avenues for improving the wavefunction expressivity and recovering parts of the remaining correlation energy.

The subsequent sections are divided into two main parts, methodology in section \ref{sec:methods} and discussion of key findings in section \ref{sec:results}. 
A theoretical overview of coupled cluster and NN-VMC electronic structure methods is contained in section \ref{subsec: theo_framework}, while \ref{subsec:BDE_form} and \ref{subsec:BDE_expt_overview} are dedicated to bond dissociation energy formalism and and an overview of existing experimental data. We provide computational details of both the explored approaches in \ref{subsec:comp_details}. In the results and discussion section, we examine the qualitative behavior of the NN-VMC wavefunctions in section \ref{subsec:NN-qualitative-comp} followed by section \ref{subsec: Extrapolation-detailed} which outlines the extrapolation methods utilized in the NN-VMC approach. Sections \ref{subsec:TE_results} and \ref{subsec:BDE_results} present and discuss the total energies and bond dissociation energies obtained for the systems considered in this study. We conclude our findings in section \ref{sec:conclusions}.

\section{Methodology}
\label{sec:methods}

\subsection{Theoretical Framework: Electronic Structure Hamiltonian and Wavefunction Approaches}
\label{subsec: theo_framework}

The stationary many-body Schr\"ondinger equation with  Born-Oppenheimer Hamiltonian that is given by
\begin{align}
{H} =
& -\sum_{i=1}^{N} \frac{1}{2}\nabla_i^2 
  -\sum_{i,A} \frac{Z_A}{|\mathbf{r}_i - \mathbf{R}_A|} \notag\\
& + \sum_{i<j}\frac{1}{|\mathbf{r}_i-\mathbf{r}_j|}
  + \sum_{A<B}\frac{Z_A Z_B}{|\mathbf{R}_A-\mathbf{R}_B|}
\label{eq:Shro}
\end{align}
where $\{ {\bf r}_i\}$ and 
$\{ {\bf R}_A,{\bf R}_B\}$ denote spatial positions of electrons and static nuclei, respectively, while $\{Z_A,Z_B\}$ are the charges.
Throughout this work, we employ pseudopotentials to replace chemically inert core electrons with an effective potential. This reduces the variational parameter space and restricts the calculation to valence energy scales, thereby improving computational efficiency and mitigating numerical and statistical errors due to lowered energy scale.
In the present study we employ correlation-consistent Effective Core Potentials (ccECPs) across both the methods, these ccECPs and the respective basis sets can be found at the Pseudopotential Library \cite{website}.
Except for the smallest systems, the exact eigenstate of this Hamiltonian, in eq ~\ref{eq:Shro} , is often inaccessible. Hence, the central challenge of the electronic structure theory is to obtain the accurate/best approximation to the exact many-electron wavefunction and/or its corresponding ground-state energy. 
While various approximate methods exist, coupled cluster methods are known for their high accuracy, and for this reason, we choose to compare our NN-based results against our coupled cluster calculations, as well as previous theory and experiment.

\paragraph{\textbf{Coupled cluster theory:}}
Coupled-cluster (CC) theory represents the correlated wavefunction as an exponential of excitation operators acting on a single reference slater determinant,
$\ket{\Psi_{CC}}=e^{\hat{T}}\ket{\Phi_0}$,
with $\hat{T}=\hat{T}_1+\hat{T}_2+\cdots$ ~\cite{cizek-1966-cc}. The exponential ansatz ensures size extensivity and incorporates disconnected excitations through the nonlinear structure of the cluster operator which enables a systematically improvable treatment of dynamic electron correlation. The amplitudes associated with $\hat{T}$ are obtained by projecting the similarity-transformed Hamiltonian onto excited determinants, giving a set of coupled nonlinear equations. Truncation at singles and doubles yields CCSD, while CCSD(T),  widely regarded as the gold standard of thermochemistry, includes a perturbative treatment of connected triple excitations. In CCSDT(Q) the theory is extended to perturbative quadruples. 
Since the theory is typically realized in a gaussian basisset, estimates of the ground state energy rely on semi-empirical, but widely validated, angular-momentum based extrapolations to the complete basis set (CBS) limit.

\paragraph{\textbf{Neural-Network Variation Monte Carlo:}}
An alternative route to \textit{ab initio} methods is provided by variational methods, where the variational ground state energy ($E_V$) is obtained by optimizing a parameterized trial wavefunction ($\psi_t$) according to the Rayleigh$-$Ritz variational principle,
\begin{equation}
E_V = \langle H \rangle=
\min_{\psi_t}
\frac{\langle \psi_t | H | \psi_t \rangle}
     {\langle \psi_t | \psi_t \rangle}.
\label{eq:variational}
\end{equation}
Due to its stochastic nature, besides the variational energy it is also instructive to evaluate the  corresponding variance (second central moment of $H$) given by 
\begin{equation}
\sigma^2 = \langle (H-E_V)^2\rangle=
\frac{\langle \psi_t | (H-E_V)^2 | \psi_t \rangle}
     {\langle \psi_t | \psi_t \rangle}.
\label{eq:variational-var}
\end{equation}
that measures energy fluctuations around $E_V$.  The variance vanishes only for an exact eigenstate and therefore serves as an additional indicator of wavefunction quality. 

Neural-network variational Monte Carlo (NN-VMC) follows this variational framework and employs highly expressive neural-network parameterizations of the many-body wavefunction. Recent developments have introduced NN ansatze in which the trial wavefunction is constructed from the slater determinants containing multi-particle permutation-equivariant orbitals, enabling highly accurate real-space descriptions of electronic correlation. The parameters of the neural network are optimized through the minimization of eq ~\ref{eq:variational}. 
Similar to conventional VMC, a common reference for NN-VMC is the antisymmetrized product wavefunction of electron orbitals $\phi(x)$,
%Hartree–Fock wavefunction, 
$\Psi (\phi(x)) = \mathcal{A}[\phi_1(x_1)\phi_2(x_2)...\phi_N(x_N)]$, where $\mathcal{A}$ denotes the antisymmetrization operator. The state of electron $i$ is denoted by $x_i=(\mathbf{r}_i,\sigma_i)$, where $\mathbf{r}_i$ and $\sigma_i$ are its spatial and spin coordinates, respectively, and $\mathbf{x}=(x_1,\ldots,x_N)$ denotes the full electronic configuration. 
In case of traditional VMC, this wavefunction contains single particle orbitals, $\Psi(\phi_j(x_i))$, which are fixed functions of individual electron coordinates and the correlation effects are often achieved through Jastrow factors \cite {jastrow-1955,kato-1957, foulkes-2001-qmcreview}. In contrast,  NN-VMC methods employ generalized many body orbitals which include explicit dependence on the full set of electronic coordinates, $\Psi(\phi_j(x_i;\mathbf{x_{\neq i}}))$.
Neural-network (NN) wavefunctions such as FermiNet, PauliNet, and Psiformer construct antisymmetric many-electron wavefunctions using determinants whose matrix elements are parameterized by neural networks~\cite{Pfau-2020-ferminet,Hermann-2020-paulinet,vonGlehn-2022-psiformer,scherbela-2025-deeperwin,li-2022-deepsolid}. Here, we focus on the Psiformer ansatz, following the notation of Ref.~\cite{vonGlehn-2022-psiformer}, the Psiformer wavefunction can be written as
\begin{equation}
\Psi_\theta(\mathbf{x}) =
\exp\!\left[J_\theta(\mathbf{x})\right]
\sum_{k=1}^{N_{\mathrm{det}}}
\det\!\left[\phi_\theta^k(\mathbf{x})\right],
\label{eq:NN_wf}
\end{equation}
where $\theta$ denotes the trainable parameters, $k$ labels the determinants, and $\Phi_\theta^k(\mathbf{x})\in\mathbb{R}^{N\times N}$ is the learned orbital matrix associated with determinant $k$. Unlike a conventional Slater determinant, whose matrix elements are single-electron orbitals, the elements of $\Phi_\theta^k(\mathbf{x})$ are permutation-equivariant functions of the full electronic configuration and can therefore encode many-electron correlations directly.
In the Psiformer, $\Phi_\theta^k(\mathbf{x})$ is constructed from electron features processed through a sequence of multihead self-attention and nonlinear layers~\cite{vonGlehn-2022-psiformer}. The self-attention mechanism allows the representation associated with each electron to incorporate information from all other electrons through learned query$-$key interactions, providing a flexible representation of electron correlation without imposing a fixed functional form for electron-electron interactions. The resulting electron representations are projected onto the elements of the determinant matrices and multiplied by exponentially decaying envelope functions to enforce the appropriate asymptotic behavior. An optional Jastrow factor $J_\theta(\mathbf{x})$ is included to impose the cusp conditions. Efficiency of other explored ansatze available in DeepQMC package, FermiNet \cite{Pfau-2020-ferminet} and PauliNet2 \cite{Hermann-2020-paulinet}, is shown in section S1 of the supplementary information.

\subsection{Bond Dissociation Energy Calculations}
\label{subsec:BDE_form}

Bond dissociation energies for each molecular species are evaluated as the energy difference between the parent molecule and its corresponding dissociation fragments. To enable a direct comparison between theoretical predictions and experimentally reported bond dissociation energies at room temperature $(D_{298}^o)$, zero-point vibrational energy (ZPE) and a finite temperature correction ($RT_{\text{correction}}$) are applied. We also include spin-orbit (SO) correction as per the system. The ZPE for the small molecules (LiH and OH) is obtained from the works of Irikura \cite{irikura-2007-ZPE}, and for TM-H we use the data provided in ref \cite{aoto-2017-icMRCC}, both these references use spectroscopic data to calculate ZPE.
Bond dissociation energy ($D_{298}^o$), to be compared with experimental results at room temperature, is calculated using the following equations,
\begin{equation}
\label{eq:BDE-eq}
D_{298}^o = D_e - \text{ZPE} + RT_{\text{correction}},
\end{equation}
where, $RT_{\text{correction}}$ = 3.7 kJ/mol~\cite{CRC}.
We obtain the equilibrium dissociation energy ($D_e$), from the energy difference of the dissociated fragments and parent molecule ($D_e^{\text{elec}}$) with added spin-orbit corrections as discussed below,
\begin{equation}
\label{eq:D_e}
D_e = D_e^{\text{elec}} + \Delta D_e^{so}
\end{equation}
The value for $D_e^{\mathrm{elec}}$ is obtained from our total energy calculations with \textit{ab initio} and NN-VMC methodologies for fragments ($E(A)$ and $E(B)$) and the parent molecule ($E(A-B)$),
\begin{align}
\label{eq:D_e_elec}
D_e^{\text{elec}} &= E(A) + E(B) - E(A-B) 
\end{align}
For systems containing Li no SO correction was required ($\Delta E_{\text{so}} = 0$) and for OH it is directly taken from ref. \cite{ruscic-2002}. Whereas, for the TM-H systems the total $\Delta D_e^{so}$ is calculated using atomic and molecular spin-orbit contributions,
\begin{align}
\label{eq:D_e_so}
\Delta D_e^{so} &= \Delta E_{\text{so}}^{\mathrm{atom}}(A) + \Delta E_{\text{so}}^{\mathrm{atom}}(B)
- \Delta E_{\text{so}}^{\mathrm{mol}}(A-B)
\end{align}
%where, $E(B) = -0.5$ Ha for hydrides.\\
Here, the molecular SO term $\Delta E_{\text{so}}^{\text{mol}}$ is from ref. \cite{aoto-2017-icMRCC}, where it was computed using CASSCF with Breit--Pauli spin--orbit coupling and for atomic SO contributions, we use the statistical average over the atomic fine structure levels \cite{aoto-2017-icMRCC},
\begin{equation}
\Delta E_{\text{so}}^{\mathrm{atom}} =
-\frac{\sum_J (2J+1),E(J)}{\sum_J (2J+1)},
\label{atomic-so}
\end{equation}
Energies $E(J)$ and total angular momentum quantum numbers $J$ are taken from the NIST Atomic Spectra Database \cite{NIST}.
It should be noted that the relativistic corrections are included naturally by employing correlation-consistent effective core potentials (ccECPs). These pseudopotentials are constructed using a tenth-order Douglas–Kroll–Hess Hamiltonian to incorporate averaged scalar relativistic effects \cite{1-ccECP,3-ccECP,4-ccECP}. 
Throughout this work, experimental bond lengths are employed, taken from the NIST database for lighter molecules (LiH and OH) and from ref ~\cite{aoto-2017-icMRCC} for transition-metal systems (TiH and NiH).
Zero-point energies (ZPE) and SO corrections ($\Delta E_{so}$) along with the respective bond lengths of the studied molecules are listed in Supplementary Table S2. 

\subsection{Experimental Bond Dissociation Energies}
\label{subsec:BDE_expt_overview}

While comparing various electronic structure methods, it is essential to benchmark theoretical results not only against each other but also against available experimental measurements and previously reported high accuracy theoretical studies in the literature. 
Experimental reference values for each molecule were collected from standard spectroscopic and thermochemical compilations, including the CRC Handbook \cite{CRC}, Lange’s Handbook \cite{langes}, and the Huber–Herzberg database \cite{huber-herzberg}. 
BDE values from these databases along with selected experimental estimates from the literature are tabulated in Supplementary Table S3.

For the lighter molecules, LiH and OH, nearly all experimental and theoretical bond dissociation energies fall within chemical accuracy of each other. The few exceptions typically correspond to the studies that either report only upper/lower bounds or rely on older experimental measurements that have since been revised. In contrast, the experimental values of bond dissociation energy reported for TiH exhibit a bimodality, clustering around two distinct values separated by approximately 50 kJ/mol, indicating unresolved experimental inconsistencies. The situation is most severe for NiH, where reported experimental bond dissociation energies show substantial scatter across the literature. This dispersion highlights the need for renewed experimental investigation and, in the interim, justifies meaningful comparison against reliable high-level theoretical predictions. 
As noted above, prior studies have been performed in service of this goal, and we add our own predictions to this body of work.

\subsection{Computational Details}
\label{subsec:comp_details}

In both, the \emph{ab initio} framework and neural-network (NN) based wavefunction approaches, we use the recently developed ccECPs, which adopt a semi-local form with averaged scalar relativistic effects and are optimized at the CCSD(T) level against all-electron atomic spectra and bonding environments. The ccECPs are accompanied by correlation-consistent basis sets constructed by minimizing the CCSD(T) ground-state total atomic energy and validated for smooth CBS extrapolation.
Details of their construction and validation are given in refs.~\cite{1-ccECP,3-ccECP,4-ccECP, gani-2020-acc-engI}, and the potentials and basis sets in various formats are publicly available in the pseudopotential library \cite{website}.
For the \textit{ab initio} calculations we employ augmented correlation-consistent atomic basis sets, (aug)-cc-p(C)V$n$Z with $n \in \{D,T,Q,5,(6)\}$, where core–valence (C)V sets are used for TM-systems. All CCSD(T) calculations are performed with  MOLPRO codebase \cite{MOLPRO}, for perturbative quadruples, CCSDT(Q), we additionally integrate MRCC package \cite{Mrcc}. 

All reported \textit{ab initio} energies are extrapolated to the complete basis set (CBS) limit. 
This CBS extrapolation is performed in a two step process, where the Hartree–Fock (HF) and correlation components are extrapolated separately according to  \cite{extrapolation}:

\begin{equation}
\label{eq:hfextrap}
    E^{\rm HF}_n = E^{\rm HF}_{\rm CBS} + a e^{- b n }
\end{equation}
\begin{equation}
%\label{eq:corrextrap}
\label{eq:cbs}
    E^{\rm corr}_{n} = E^{\rm corr}_{\rm CBS} + \frac{\alpha}{(n+3/8)^{3}} + \frac{\beta}{(n+3/8)^5}
\end{equation}
Here, $n$ is the cardinal number of the basis set and $E^{\rm HF}_{CBS}$ and $E^{\rm corr}_{CBS}$ denote the extrapolated HF and correlation energies, respectively. 
Direct CCSDT(Q) calculations with larger basis sets (QZ--5Z) were computationally prohibitive, therefore, the corresponding total energies are estimated using the previously validated ratio-based approach described in Ref.~\cite{aqsa-2025-acc-eng-II}. Total energies at various basis levels and the corresponding extrapolated values for all the studied systems are listed in Supplementary Table S4.

Moving on to the second electronic structure method, all the neural-network VMC (NN-VMC) calculations are performed using the DeepQMC package \cite{deepqmc}, which utilizes JAX \cite{2018-jax} for efficient execution on accelerator hardware. As discussed in section ~\ref{subsec: theo_framework}, due to its enhanced ability to capture electronic correlation, all quantitative comparisons of BDEs presented in this work are performed using the Psiformer ansatz \cite{vonGlehn-2022-psiformer}, with default parameters provided by DeepQMC package, unless specified otherwise. 
During the training phase, VMC optimization was carried out using the Kronecker-Factored Approximate Curvature (KFAC) \cite{2015-kfac} method in combination with a hybrid electronic sampler. 
A walker batch size of 4096 is used with $10^5$ training iterations performed for each atom and molecule. 
The overall computational workflow is divided into training and inference phases, both of which were carried out on individual compute nodes containing 4 NVIDIA A100 GPUs.
Details of the computational expense with a summary of the computational setup and parameters is provided in supplementary section S4 and Table S5, respectively.

To estimate the asymptotic energies within practical computational budgets, we employ two post-training extrapolation schemes: zero-variance ($\sigma^2 \to 0$) and infinite-step  ($t \to \infty$) extrapolations. These procedures mitigate residual optimization bias arising from finite training length. Details of the extrapolation protocols that we have explored and further developed, along with the resulting energies, are discussed in sections \ref{subsec: Extrapolation-detailed} and \ref{subsec:TE_results} respectively.

\section{Results and discussion}
\label{sec:results}
\subsection{NN-VMC Training Efficiency}
\label{subsec:NN-qualitative-comp}

Before moving on to the resulting energies for the studied systems, first we assess the accuracy of our neural-network implementation by comparing our training efficiency with previously published results for the Ti atom which was obtained using the FermiNet ansatz combined with ccECPs \cite{li-2022-ccECP-ML}. Here we employ the same training parameters as provided by Li.\textit{ et. al} in ref. \cite{li-2022-ccECP-ML}.
Notably, the Psiformer ansatz in DeepQMC, using default hyperparameters, recovers a comparable fraction of the correlation energy with approximately five times fewer training steps than reported by Li.\textit{ et. al}, 
%\aqsa{should we give exact numbers?}, 
suggesting the improved efficiency of the attention based architecture as compared to the published FermiNet based results.
We also observe that the DeepQMC implementation of the FermiNet ansatz yields variational energies similar to those reported in Ref.~\cite{li-2022-ccECP-ML}.

\begin{figure}[!htpb]
    \centering
    \includegraphics[width=1.0\linewidth]{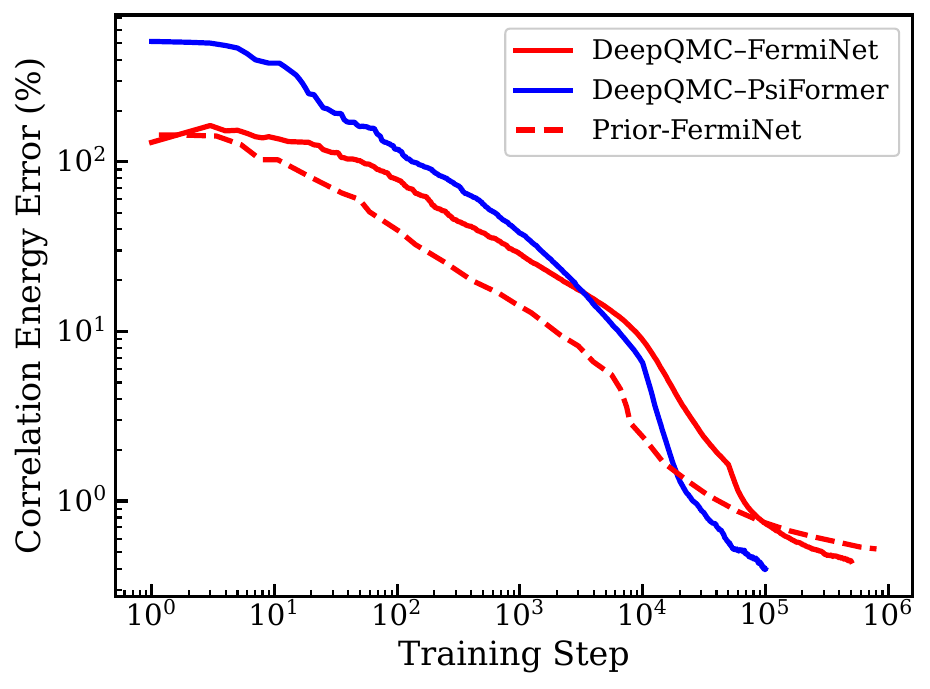}
    \caption{
    Training curves for the Ti atom, shown as the percentage of residual correlation energy with respect to CCSDT(Q)/CBS results. DeepQMC based optimization with FermiNet architecture (solid red) is qualitatively similar to the FermiNet based result in ref. ~\cite{li-2022-ccECP-ML} (dashed red).  Notably, the Psiformer ansatz \cite{vonGlehn-2022-psiformer} (blue) recovers a greater amount of correlation energy in fewer training steps. Median energy over the last 10\% of iterations is shown for clarity.
    }
    \label{fig:ccECP-ML-paper}
\end{figure}

\begin{figure*}[!htpb]
    \centering

    \begin{subfigure}{0.49\textwidth}
        \centering
        \includegraphics[width=\linewidth]{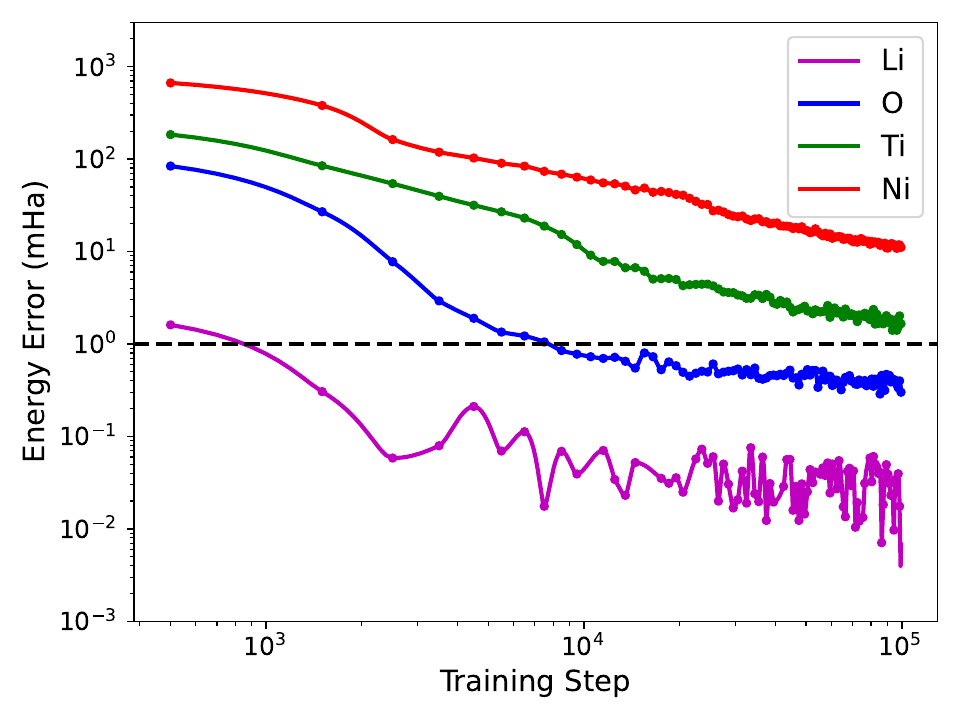}
        \caption{}
        \label{fig:psi-energies}
    \end{subfigure}
    \hfill
    \begin{subfigure}{0.49\textwidth}
        \centering
        \includegraphics[width=\linewidth]{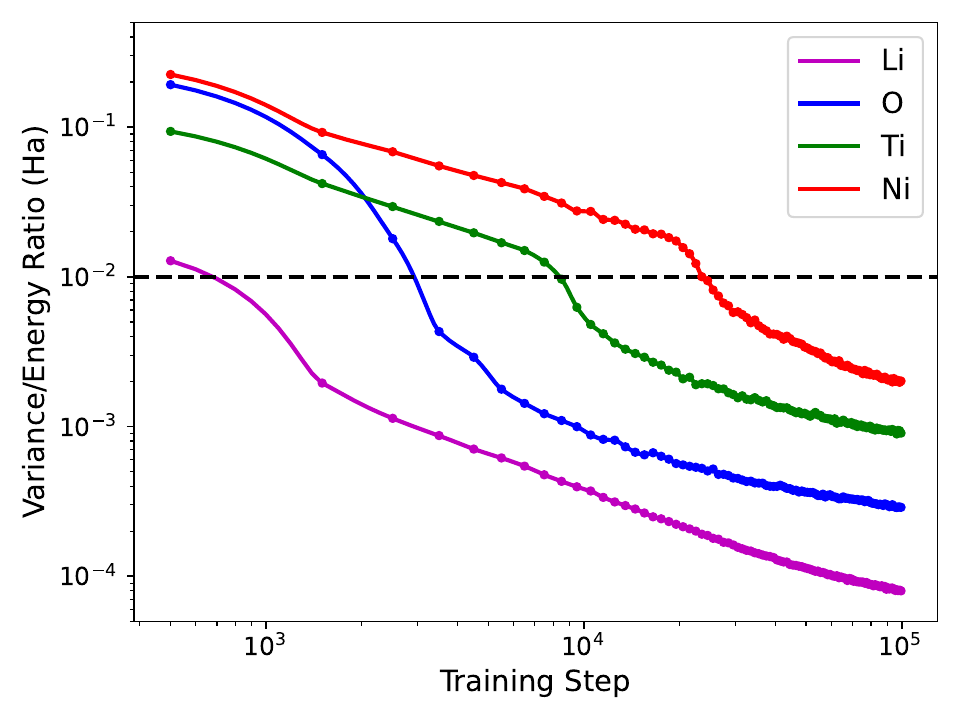}
        \caption{}
        \label{fig:psi-ve-ratio}
    \end{subfigure}

    \caption{Convergence behavior of the Psiformer ansatz (16 determinants) during training. 
    \textbf{(a)} Total energy errors for Li, O, Ti, and Ni atoms with respect to the CCSDT(Q)/CBS reference energies. Accuracy at the 1 mHa level is denoted by the horizontal dashed line. 
    \textbf{(b)} Variance-to-energy ratio, the horizontal dashed line corresponds to a level not often crossed in the traditional Slater-Jastrow ansatz (0.01 Ha) \cite{qmcpack_manual}.
    }
    \label{fig:psi-convergence}
\end{figure*}

To further assess the performance of the Psiformer architecture, 
we examine the convergence curves across the atomic systems considered in this study: Li, O, Ti, and Ni. Figures~\ref{fig:psi-energies} and \ref{fig:psi-ve-ratio} outline the training behavior of the Psiformer ansatz in terms of the total energy error and the variance-to-energy ratio ($\sigma_V^2/\lvert E \rvert $), which together provide complementary indicators of convergence. The curves are constructed by evaluating the wavefunction at a series of saved training checkpoints through separate inference calculations. The total energy error, in Figure ~\ref{fig:psi-energies}, is relative to the calculated CCSDT(Q)/CBS energies for the respective atoms. \aqcom{get the plotting script of ~\ref{fig:psi-energies} from Jaron to ensure CCSDT(Q)/CBS values used are up-to-date.}
For all systems, the training initially shows a rapid decrease in energy error followed by a slower asymptotic regime. 
The convergence rate depends on the atomic number, e.g. lighter systems such as Li and O reach chemical accuracy ($<$1 kcal/mol) with respect to CCSDT(Q)/CBS earlier in training, whereas the heavier TM-systems require longer optimization. By the end of our training, using the set up listed in supplementary Table S5, Ni remains approximately 12 mHa above the CCSDT(Q)/CBS reference, while the other atoms demonstrate that the neural-network wavefunction is capable of reaching close to high-accuracy theoretical values.
Moving to the quality of wavefunction demonstrated in Figure ~\ref{fig:psi-ve-ratio}, a substantial improvement in the  V-E-ratio, $\sigma_V^2/\lvert E \rvert$, relative to the conventional Slater-Jastrow ansatz is observed \cite{ve-ratio-LiNiO-Jaron-2026, qmcpack_manual}. Remarkably, even for Ni, which exhibits the slowest convergence among the atomic systems studied, the V-E-ratio drops to approximately 0.01 Ha within the first quarter of the training process, highlighting the effectiveness of the ansatz in rapidly approaching a high-quality wavefunction.

\subsection{NN-VMC Extrapolation Protocols}
\label{subsec: Extrapolation-detailed}

In an attempt to arrive at the most accurate estimates possible for atomic/hydride total and binding energies, we employ two distinct extrapolation approaches.  The first has been applied by the QMC community for a long time: energy vs energy variance extrapolation - $(E,\sigma^2)$ \cite{mitas-1991,ceperly-1993-2d, mitas-1993}. The second is newer, based on observations of the progression of the NN-VMC training energies, where power-law behavior is evident - $(E,1/t^\alpha)$ \cite{scherbela-2025-inv-step}.
In each case, linearity between the primary variables is assumed:
\begin{align}
    E_{\sigma^2}  &= E_{\sigma^2 \to 0} + m\sigma^2 
    \label{eq:ext-eq-var}\\
    E_{t} &= E_{t \to \infty} + m/t^\alpha
    \label{eq:ext-eq}
\end{align}
Where, $E_{\sigma^2 \to 0}$ and $E_{t \to \infty}$ represent the extrapolated asymptotic energies for variance ($\sigma^2$) and training step ($t$) extrapolations and $m$ is the slope of the fit. In keeping with this assumption, we use a robust regression scheme to extract essential linear relationships in the data.  The Theil-Sen \cite{theil-1950, sen-1968} method is a known robust regression approach.  We further reduce outlier sensitivity by using RANSAC \cite{ransac-1981}, with Theil-Sen as its interior regressor. 
The combined systematic and statistical uncertainty is estimated in the following way.
We draw a large number of 50/50 split samples from the $(E,\sigma^2,t)$ data points (sub-sampling) to represent systematic variability in the points generated by the training process. 
For each 50/50 split, we include statistical error via resampling based on the estimated error on the mean energy at each training point. For each of these perturbed sub-datasets we perform robust linear regression as just described. 
We then extract the mean estimates of the asymptotic energies ($E_{\sigma^2 \to 0}$ and $E_{t \to 0}$) and their uncertainties from the distribution of extrapolated values.

\begin{figure}[!htpb]
    \centering
    \includegraphics[width=1.0\linewidth]{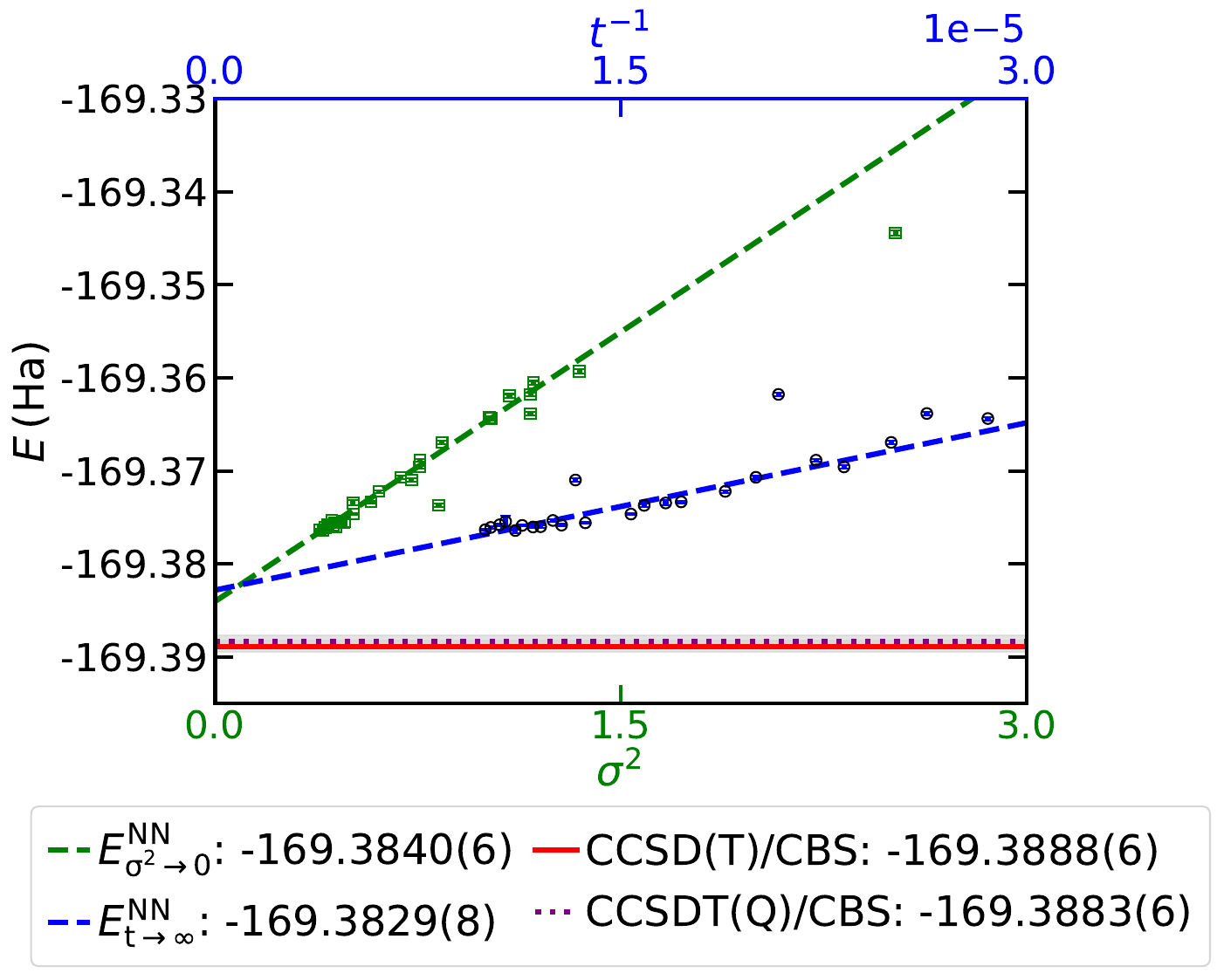}
    \caption{
    Extrapolation of the NN-VMC total energy toward zero variance ($E_{\sigma^2 \to 0}$) (green) and infinite training steps ($E_{t \to \infty}$) (blue) shown for the Ni atom with 16 determinants in the NN-wavefunction ansatz.
    Robust linear fitting is performed via the methods described in \ref{subsec: Extrapolation-detailed}.
    The horizontal lines refer to CBS extrapolated CCSD(T) and CCSDT(Q) energies.
    }
    \label{fig:Ni-extrap}
\end{figure}

The bare training data are often extremely noisy, producing a large number of extreme outliers.
To remedy this, we consider two different approaches to obtain well-behaved data for subsequent regression.  In the first approach, we notice that the intermediately optimized wavefunctions show 
reduced sensitivity relative to the training energies produced at low sampling. 
We therefore performed inference calculations on an evenly spaced subset of the training steps. 
Since these inference calculations are performed with substantially larger sampling than the corresponding training evaluations, they possess significantly smaller statistical uncertainties and provide a more reliable basis for the extrapolation.
As an alternative, for inverse step extrapolation only, we also perform smoothing on the original training data without additional inference. 
To smooth the data, we obtain the median energy over fixed ranges (1000 steps) of the training curves. 
This is beneficial because the median is quite insensitive to outliers. One example of energy variance and step based extrapolations is shown in Figure ~\ref{fig:Ni-extrap} for the Ni atom (see SI for the rest of the systems). Energy variance extrapolation is shown in green and step-based extrapolation in blue. 
The horizontal lines beneath the linear fits are the total energies obtained from basisset extrapolated CCSD(T) and CCSDT(Q). 
While in this example the two extrapolated NN-VMC values closely agree, approximately 1$\sigma$ apart, this need not be the case in general. 
Indeed, we observe small differences in a number of cases, as reported in Table \ref{tab:energy_bind} (also see supplementary Figures S3, S6, S9 and S12).
We take these differences as a final indicator of uncertainty (or the lack thereof) in our predicted ground state energies based on the Psiformer wavefunction ansatz.

\subsubsection{Extrapolations Based on Energy Variance}

In quantum Monte Carlo studies, extrapolations based on ratios of the first and second moment of the Hamiltonian have been used for some time, see for example \cite{mitas-1991,mitas-1993,ceperly-1993-2d}. More recently, NN based study, by W. Fu \textit{et al}, have also found variance based extrapolation to provide a reliable estimate of the asymptotic energies. \cite{fu-ji-chen-2024-var-ext}.
As we now demonstrate, an approximate scaling relation that provides the basis for data analysis can be placed on a more sound mathematical footing in the limit of high quality wavefunctions. 
Neural-network wavefunctions certainly fall within this class, as many prior studies in addition to our own have shown that it could obtain correlations at the $\sim$ 99\% level and beyond (shown in Figure \ref{fig:ccECP-ML-paper} and later in Figures \ref{fig:cor_Ti_TiH} and \ref{fig:cor_Ni_NiH}).

By expanding a given trial wavefunction in the basis of the many-body eigenstates, $\lvert \Psi_t \rangle =c_0 \lvert \Psi_0 \rangle+\sum_{i>0}{c_i} \lvert \Psi_i \rangle$ with energies $E_i$, we can define the \emph{contamination} in the first and second moments of the energy distribution as
\begin{align}
    E_S &= \frac{\langle H\rangle-E_0}{1-c_0^2} \\
    E_D^{(2)} &= \frac{\langle H^2\rangle-E_0^2}{1-c_0^2}-2E_0E_S
\end{align}
where $\langle H^n\rangle=\langle\Psi_t\lvert H^n\lvert\Psi_t\rangle$.  
Using these expressions for $\langle H\rangle$ and 
$\langle H^2\rangle $ one can derive the following (exact) identity
\begin{align}
 \langle H\rangle= E_0+
 {E_S\over E_D^{(2)}} [\sigma_V^2+(\langle H\rangle -E_0)^2]
 \label{eq:mom2}
 \end{align}
 For accurate trial functions we have 
 $\sigma_V^2 \gg (\langle H\rangle -E_0)^2$
 and we can write 
 \begin{align}
 \langle H\rangle= E_0+
 {E_S\over E_D^{(2)}} \sigma_V^2+{\cal{O}}[(\langle H\rangle-E_0)^2]
 \end{align}
 The proportionality factor is the ratio of the contaminations that are both positive.
 More details regarding the derivation can be found in supplementary section S5.
The last expression shows that for accurate trial wavefunctions the dominant scaling of the total energy is nominally proportional to the variance. However, there remain important issues to consider. 
One issue is that  the proportionality constant in front of $\sigma_V^2$, in fact,  still {\em indirectly} depends on $\sigma_V^2$, ie, the perfect linearity is not guaranteed even when the explicitly quadratic error term is small. Empirically, it has turned out that this indirect dependence might be visible and depending on the details of the wavefunction this effect can be pronounced or diminished.
We also have to consider that the dependence $\langle H\rangle$ vs $\sigma_V^2$ can exhibit irregularities (eg, lower variational energy can lead to higher variance and vice versa) that might occur during training and therefore one could expect some additional scatter from inherent VMC randomness. 
Interestingly, indeed even high accuracy variational results show some dispersion. However, we were able to reach meaningful extrapolations that has enabled us to either confirm agreement with independent, essentially accurate CC/CBS calculations, or to point out residual level of variational bias as shown in the following sections. 

\subsubsection{Extrapolations Based on Training Step}
\label{subsubsec:step-extrap}

A complementary and independent predictor of total energies is extrapolation based on the training steps ($t$). 
For this extrapolation, we employ the empirical power law approach proposed by Scherbela \textit{et al.} \cite{scherbela-2025-inv-step}, in which the residual error decays as $ E(t)-E_{t \to \infty} \propto t^{-\alpha}$, with $\alpha \approx 1$ and $t$ is the training/optimization step. 
Here we explicitly consider how the resulting estimates of $E_{t \to \infty}$  depend on the choice of exponent ($\alpha$) and the portion of the training trajectory included in the fit.

\begin{figure}[!htpb]
    \centering
    \includegraphics[width=1.0\linewidth]{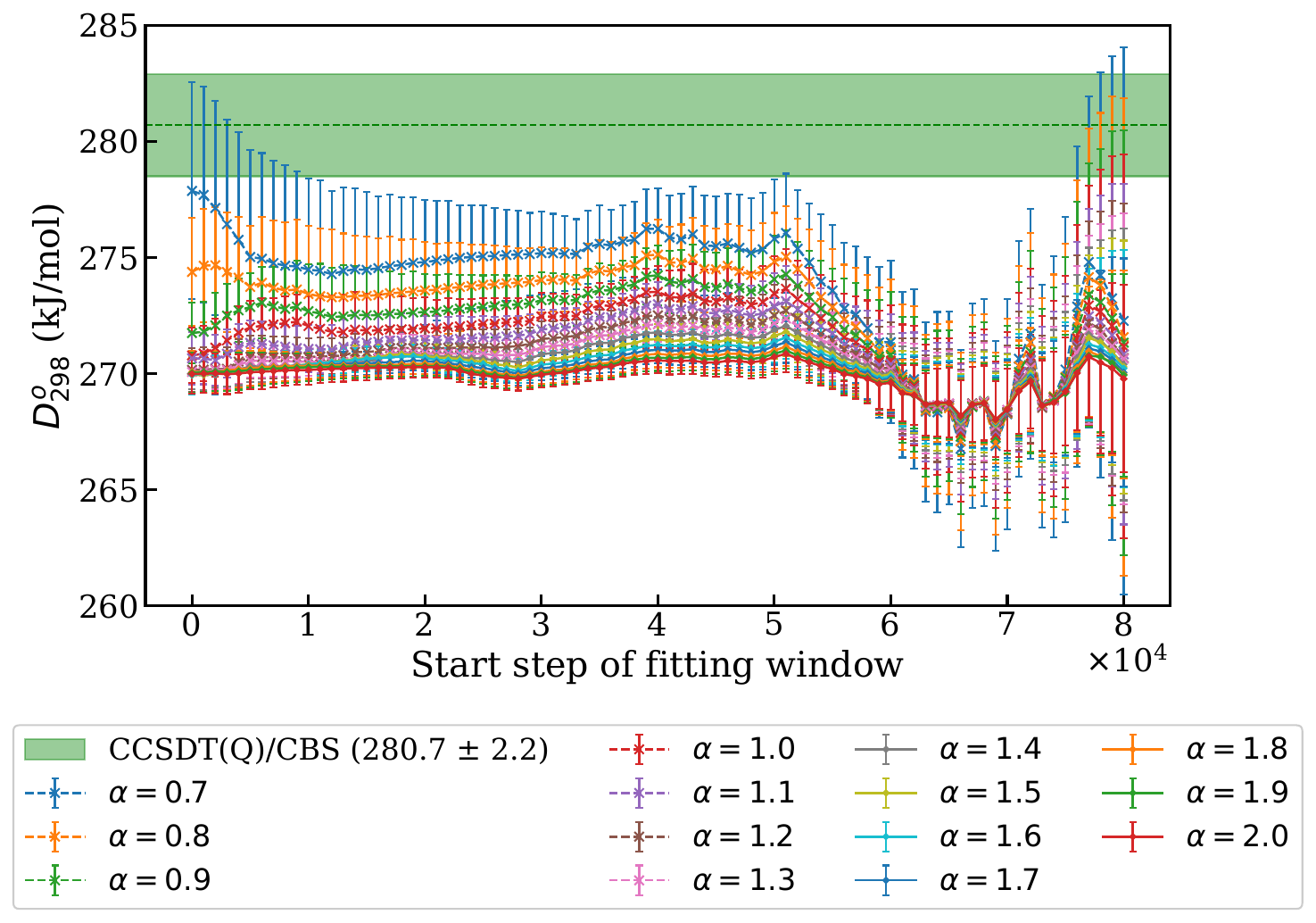}
    \caption{Sensitivity of the infinite step extrapolated $D_{298}^o$(NiH) to the choice of the polynomial decay factor ($\alpha$) and the fitting window. Each point on the x-axis specifies the beginning of the fitting window, such that the corresponding fit is performed over the interval ([x,EOT]), where EOT denotes the end of training.
    }
    \label{fig:NiH-BDE-train-smooth-alpha}
\end{figure}

The first set of investigation is performed on median smoothed training data, in the absence of large number of inference points one can utilize training trajectory to estimate the asymptotic energy. We perform multiple fits for each value of $\alpha$ ($\alpha\in[0.7,2.0]$) as successively larger amounts of the early training data are excluded.  
Whenever proportionality to $1/t^\alpha$ holds well, we expect a stable prediction of the energy. 
As a demonstration, we choose NiH binding energies here since Ni systems are least converged in total energy with respect to the finite training and thus represent the largest challenge for extrapolation.
As shown in Figure \ref{fig:NiH-BDE-train-smooth-alpha} for NiH, we find that over a significant range of excluded data, the extrapolated energies converge to a plateau separately for each value of $\alpha$ which represent the best estimates.
We also observe that the inclusion of early training steps results in larger $\alpha$ dependence, due to abundant outliers in initial training stage, suggesting that the asymptotic convergence regime has not yet been reached. 
For example, while quantifying the $\alpha$ sensitivity within each fitting window relative to $\alpha=1$, the decay exponent proposed in ref.~\cite{scherbela-2025-inv-step}, the largest deviation in the mean $D_{298}^o$ is 7 kJ/mol, obtained for $\alpha=0.7$ when fitting the full training trajectory. Excluding the initial $ 2 \times 10^4$ steps reduces this deviation to below $\sim3$ kJ/mol. This is an expected behavior due to the fit running over two distinct training phases; early transient dynamics and late asymptotic region.
Further on, when more than half the training steps have been excluded, the fitted data gain higher variability with higher statistical uncertainty, as shown in Figure ~\ref{fig:NiH-BDE-train-smooth-alpha}, indicating that robust extrapolation requires retaining data from the intermediate training regime through the end of training (EOT). Together, these results show that the extrapolation is most sensitive to the inclusion of initial training regime or excluding too large a fraction of the trajectory.
The same qualitative dependence on the fitting window and $\alpha$ is reproduced when the analysis is performed using inference checkpoints for the asymptotic energy estimates of other studied systems (supplementary Figures S7, S10, and S13). Based on these observations we exclude the initial $2\times10^4$ training steps in all final energy  extrapolations and adopt $\alpha =1$, as also observed by ~\cite{scherbela-2025-inv-step}. For these parameters, we obtain $D_{298}^o$ (NiH) =\BENiHstepmoredet ~kJ/mol.

\begin{table*}[!htpb]
\centering
\caption{Total energies (Ha) and BDEs, $D_{298}^o$ kJ/mol (eq ~\ref{eq:BDE-eq}), for TiH and NiH. Listed are our results (NN-VMC, CCSD(T) and CCSDT(Q)), prior theoretical and experimental estimates. Where relevant, extrapolation methods used to arrive at the final results are specified. EOT (end-of-training) refers to the energies obtained by performing inference calculation at the last trained checkpoint (after training for $10^5$ optimization steps). For TiH, dataset of inference points along the training curves is used for extrapolation whereas NiH uses median-smoothed training data for extrapolation as described in \ref{subsubsec:step-extrap}. For NiH, here we show energies obtained from expanded wavefunction ansatz with 160 determinants, energetics via default 16 determinant ansatz can be found in
Figures S12 and S14 in supplementary information.
}
\label{tab:energy_bind}
\begin{tabular}{|l|lllll|}

\hline
System & Method & Extrap. & $E_\mathrm{atom}$ (Ha) & $E_\mathrm{mol}$ (Ha) & $D_{298}^o$ (kJ/mol) \\
%\midrule
\hline

\rule{0pt}{3ex} & \textbf{This Study:} & & & & \\
\textbf{~~TiH} & NN-VMC    & EOT      &\Tieot    &\TiHeot    &\BETiHeot    \\
 & NN-VMC       & eq ~\ref{eq:ext-eq} ~($t \to \infty$)  ~    &\Tistep    &\TiHstep    &\BETiHstep    \\
 & NN-VMC       & eq ~\ref{eq:ext-eq-var} ~($\sigma^2 \to 0$) ~    &\Tivar     &\TiHvar     &\BETiHvar     \\
 & CCSD(T)-QZ   &                &\Ticctqz   &\TiHcctqz   &\BETiHcctqz   \\
 & CCSD(T)-5Z   &                &\Ticctfivez&\TiHcctfivez&\BETiHcctfivez\\
 & CCSD(T)-CBS  &~eqs ~\ref{eq:hfextrap}~-~\ref{eq:cbs}~ &\Ticctcbs  &\TiHcctcbs  &\BETiHcctcbs  \\
 & CCSDT(Q)-CBS &~eqs ~\ref{eq:hfextrap}~-~\ref{eq:cbs}~&\Ticcqcbs  & \TiHccqcbs &\BETiHccqcbs  \\
\cline{2-6}

\rule{0pt}{3ex} & \textbf{Prior Theory:}    &       &    &    &    \\
 & CCSD(T)-QZ\cite{moltved-2019-ccsdt}    &               &  &  &  210.2      \\
 & CCSD(T)(CV)/CBS\cite{aoto-2017-icMRCC}           &               &  &  &  200.2      \\
 & CCSD(T)(CV)/CBS+$\Delta_\mathrm{MR}$\cite{aoto-2017-icMRCC} &               &  &  &  215.7      \\
 & AFQMC/TZ\cite{Shee-2019-AFQMC}         & CCSD(T) corr. &  &  &  194.8(3.8) \\
\cline{2-6}

\rule{0pt}{3ex} & \textbf{Prior Expt.:}    &       &    &    &    \\
 & CRC \cite{CRC}              &               &  &  &  \tihcrc \\
 & Armentrout \cite{armentrout-1996-TiH-NiH}&               &  &  &  192.7(6.0) \\

%\cline{1-5}
\hline
\hline
\rule{0pt}{3ex} & \textbf{This Study:} & & & & \\
\textbf{~~NiH} & NN-VMC   & EOT       & \Nieotmoredet    &\NiHeotmoredet    & \BENiHeotmoredet    \\
& NN-VMC   & eq ~\ref{eq:ext-eq} ~($t \to \infty$)  ~       & \Nistepmoredet    &\NiHstepmoredet    & \BENiHstepmoredet    \\
% & NN-VMC      & zero var.       & \Nivar     &\NiHvar     & \BENiHvar     \\
 & CCSD(T)-QZ  &                 & \Nicctqz   &\NiHcctqz   & \BENiHcctqz   \\
 & CCSD(T)-5Z  &                 & \Nicctfivez&\NiHcctfivez& \BENiHcctfivez\\
 & CCSD(T)-CBS & ~eqs ~\ref{eq:hfextrap}~-~\ref{eq:cbs}~& \Nicctcbs  &\NiHcctcbs  & \BENiHcctcbs  \\
 & CCSDT(Q)-CBS& ~eqs ~\ref{eq:hfextrap}~-~\ref{eq:cbs}~& \Niccqcbs  &\NiHccqcbs  & \BENiHccqcbs  \\
\cline{2-6}
\rule{0pt}{3ex}  & \textbf{Prior Theory:}    &       &    &    &    \\
 & CCSD(T)-QZ\cite{moltved-2019-ccsdt}       &               &  &  &  293.1      \\
 & CCSD(T)(CV)/CBS   \cite{aoto-2017-icMRCC}           &               &  &  &  275.2      \\
 & CCSD(T)(CV)/CBS+$\Delta_\mathrm{MR}$   \cite{aoto-2017-icMRCC}&               &  &  &  290.7      \\
 & AFQMC/TZ \cite{Shee-2019-AFQMC}           & CCSD(T) corr. &  &  &  270.2(4.6) \\
 & AFQMC/CBS\cite{Shee-2019-AFQMC}           & TZ--QZ basis sets       &  &  &  248.2(7.4) \\
\cline{2-6}
\rule{0pt}{3ex}  & \textbf{Prior Expt.:}    &       &    &    &    \\
 & CRC      \cite{CRC}           &               &  &  &  \nihcrc     \\
 & Tolbert \cite{tolbert-1986-NiH}   &               &  &  &  272(25)    \\
\hline
%\bottomrule
\end{tabular}
\vspace{2pt}
\end{table*}

%%~~~~~~~~~~~~~~~~~~~~~~~~~~~~~~~~~~~~~~~~~~~~~~~~~~~~~~~~~~~

\subsection{Total Energy Results}
\label{subsec:TE_results}

\paragraph{\textbf{Li and O containing systems:}}

Supplementary Tables S4 and S6 report the total energies entering eq ~\ref{eq:D_e_elec} for these \textit{sp-}systems. These total energies were computed using two complementary methodologies: conventional \textit{ab initio} quantum chemistry methods and neural network based Variational Monte Carlo (NN-VMC). For the smallest systems in this study, atomic Li and LiH, calculations were carried out using both ccECP-based pseudopotentials and all-electron Hamiltonians. Removal of the $1s^2$ core reduces Li and LiH to effective one-electron and two-electron systems for which HF/CBS and CISD/CBS values are reported as exact within a given basis set. For the all-electron case we provide the total energies obtained with FCI/CBS. 
The NN-VMC energies, both at the end of training and after extrapolation, agree within \textit{ab initio} CBS energies within their respective uncertainties (see supplementary Table 6).

For atomic oxygen and its hydride, ccECP-based \textit{ab initio} and stochastic NN-VMC calculations were performed where we find the EOT and extrapolated asymptotic NN-VMC energies to be within sub-milli Ha agreement with the CBS-extrapolated CCSD(T) results as shown in supplementary Table S6. Inclusion of higher-order correlation through CCSDT(Q)/CBS further lowers the total energies of both O and OH by approximately $0.5$ mHa. For the oxygen atom, however this correction remains within the combined uncertainties and is therefore indistinguishable from the CCSD(T)/CBS value. 
For these systems, the agreement of NN-VMC with the reference is reached purely at the variational level and no extrapolations were necessary. More detailed discussion of energetics involving these prototypical \textit{sp}-system can be found in supplementary section S7.

\paragraph{\textbf{Ti and TiH:}}

\begin{figure}[!htpb]
    \centering
    \includegraphics[width=1.0\linewidth]{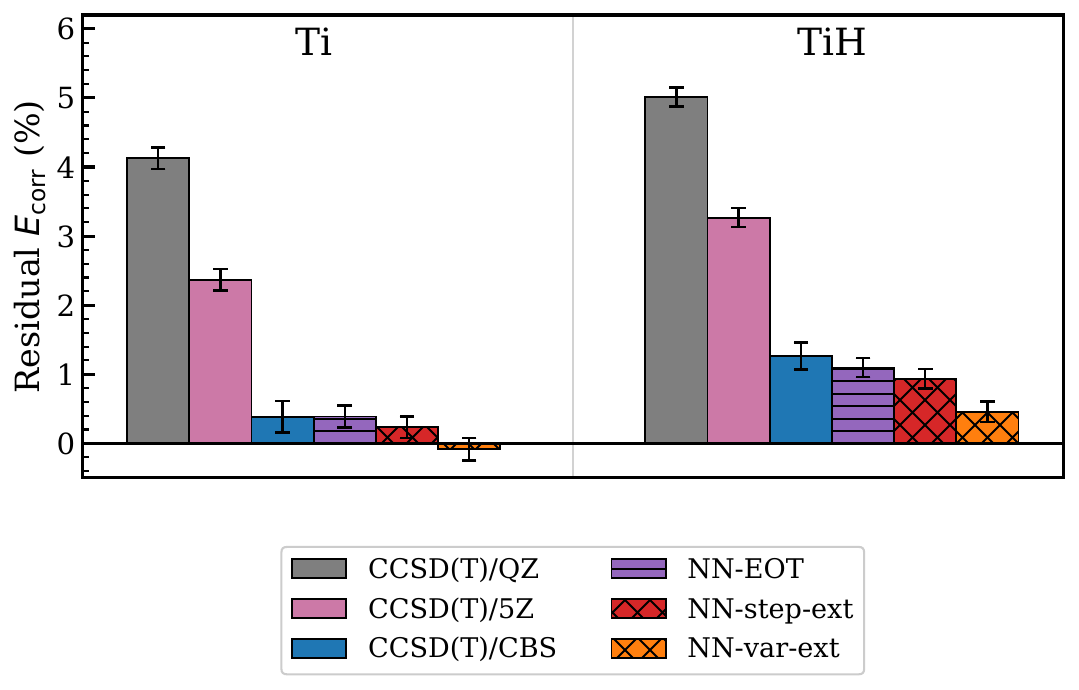}
    \caption{
    Percentage of uncaptured correlation energy for Ti and TiH through the two studied eletronic structure methods. 1. CCSD(T) at QZ and 5Z basis levels and in the CBS extrapolated limit. 2. NN-VMC at end of training (EOT), after extrapolation through infinite-step (NN-step-ext) and zero-variance (NN-var-ext) approaches. In each case, CCSDT(Q)/CBS is used as the reference to estimate the full correlation energy and the error bars show the combined propagated uncertainty of each respective method and the reference.
    }
    \label{fig:cor_Ti_TiH}
\end{figure}

For the early transition-metal systems, Ti and TiH, using the Psiformer ansatz with the parameters listed in supplementary Table S5, we observe good performance for both the lowest achieved variational energies and those including extrapolation.
Within all comparisons, whether CCSD(T) or NN-VMC, the molecular system is substantially more difficult, perhaps due to symmetry lowering intrinsic to bond formation.
In Figure ~\ref{fig:cor_Ti_TiH}, we compare NN-VMC with CCSD(T) against the CCSDT(Q)/CBS benchmark. The quantity shown is the percentage of missing correlation energy. For CCSD(T), the finite but large QZ and 5Z basis sets miss more than 4\% and 2\% of the correlation energy and in the complete basis set limit reaching or exceeding $\sim$1\% accuracy with respect to CCSDT(Q)/CBS.  Our final variational energies obtained at the end of $10^5$ training steps (NN-EOT) already matches the quality of CCSD(T)/CBS.
The infinite-step extrapolated NN-VMC energies show only a small increase in gained correlation energy, while the zero-variance extrapolation reduces the error to 0.5\% or less. 
Overall, NN-VMC with the Psiformer ansatz shows excellent quality as a benchmark level electronic structure methods for the Ti-based systems.

\paragraph{\textbf{Ni and NiH:}}

\begin{figure}[!htpb]
    \centering
    \includegraphics[width=1.0\linewidth]{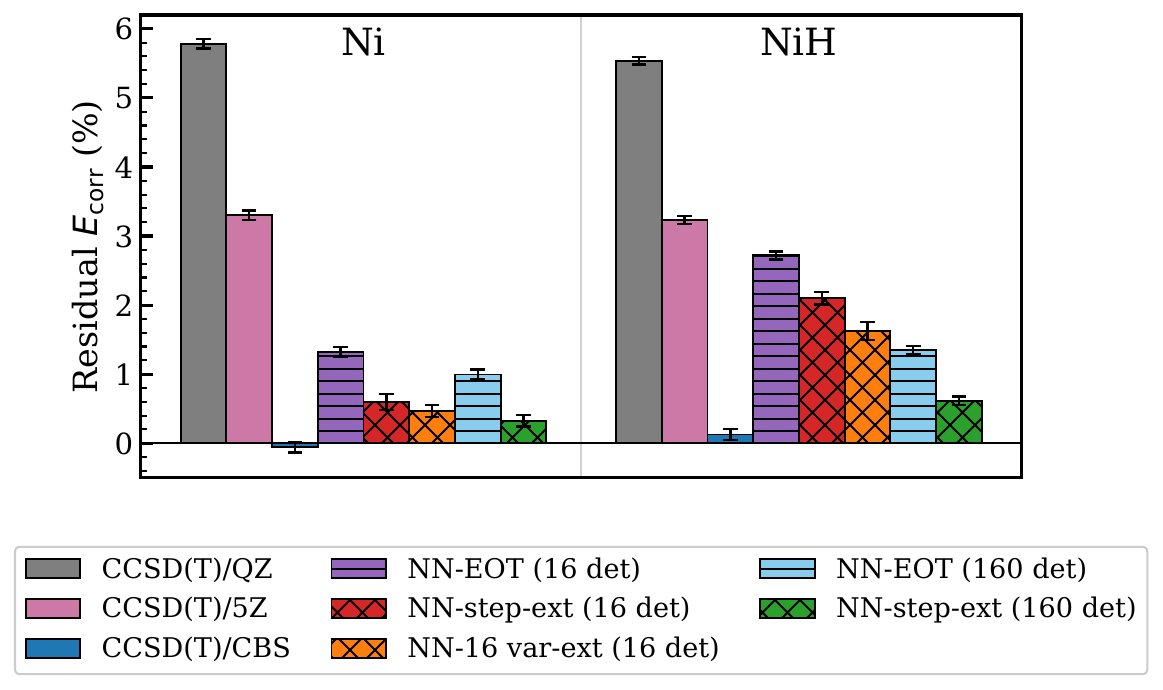}
    \caption{
    Same as Figure ~\ref{fig:cor_Ti_TiH}, now for Ni and NiH, with NN-VMC results shown for both 16 and 160 determinant Psiformer ansatze.
    }
    \label{fig:cor_Ni_NiH}
\end{figure}

The situation is different for the most challenging systems considered here, namely Ni and NiH. 
Figure~\ref{fig:cor_Ni_NiH} similarly shows the remaining correlation energy relative to the CCSDT(Q)/CBS benchmark. 
Somewhat surprisingly, in contrast to Ti/TiH, in these cases CCSD(T)/CBS shows complete agreement with CCSDT(Q), while NN-VMC shows increased error.
Considering the case of default Psiformer ansatz with 16 determinants (NN-16), NN-VMC still recovers more correlation energy at the end of training ($10^5$ training steps) than the \textit{finite-basis} CCSD(T)/5Z but remains approximately 12 and 28 mHa above the CCSDT(Q)/CBS benchmarks for atomic and hydride systems, respectively. 

We should also note that at fixed training length ($10^5$), NN-VMC reaches less of its full potential in each respective system from LiH to NiH.  This is illustrated particularly well in Figure ~\ref{fig:psi-ve-ratio}, where the independent variance-to-energy ratio metric shows reduced quality in this exact ordering, while also suggesting that equal quality might be reached between the systems at successively larger training times at higher electron count.
To determine whether the remaining difference could be attributed to incomplete optimization, we extended the NiH training to $2 \times 10^5$ steps followed by a final inference calculation (see supplementary Figure S2(a)).
Despite the doubled training time, corresponding energy improvement at the end of extended training remained limited to $\sim$ 2 mHa.
This modest gain suggests that the optimization trajectory is approaching saturation, hence, further improvement likely requires a more expressive representation of the wavefunction.
Given the well-established multireference character of NiH \cite{jiang-2012-niH-multi-ref} and the near-degeneracies present in the Ni electronic manifold, the determinant content of the ansatz is expected to play an important role. To investigate this effect in case of NiH, which remained farthest from the CCSDT(Q)/CBS reference, we increased the number of determinants in the Psiformer ansatz from the default value of 16 to 160 and the further recovered correlation energy (in percentage) is shown in Figure ~\ref{fig:cor_Ni_NiH} (NN-160).
The resulting infinite step extrapolated value (using median-smoothed training curve) improves by $\sim$ 15 mHa over the 16 determinant case, reducing the difference from the CCSDT(Q)/CBS reference from 22 mHa to 6 mHa -- a 3.5x improvement -- which brings the the level of missing correlation energy down to $<1$\% similar to other cases. A comparison of training curves obtained from 16 and 160 determinant Psiformer anstaz is shown is supplementary Figure S2(b). Importantly, a tenfold increase in determinant count results in only about a twofold increase in the computational cost. This suggests the ultimate limit of the Psiformer ansatz promises much greater gains and merits further systematic investigation in later studies.

\subsection{Bond Dissociation Energies}
\label{subsec:BDE_results}

\begin{figure*}
    \centering
    \includegraphics[width=1\linewidth]{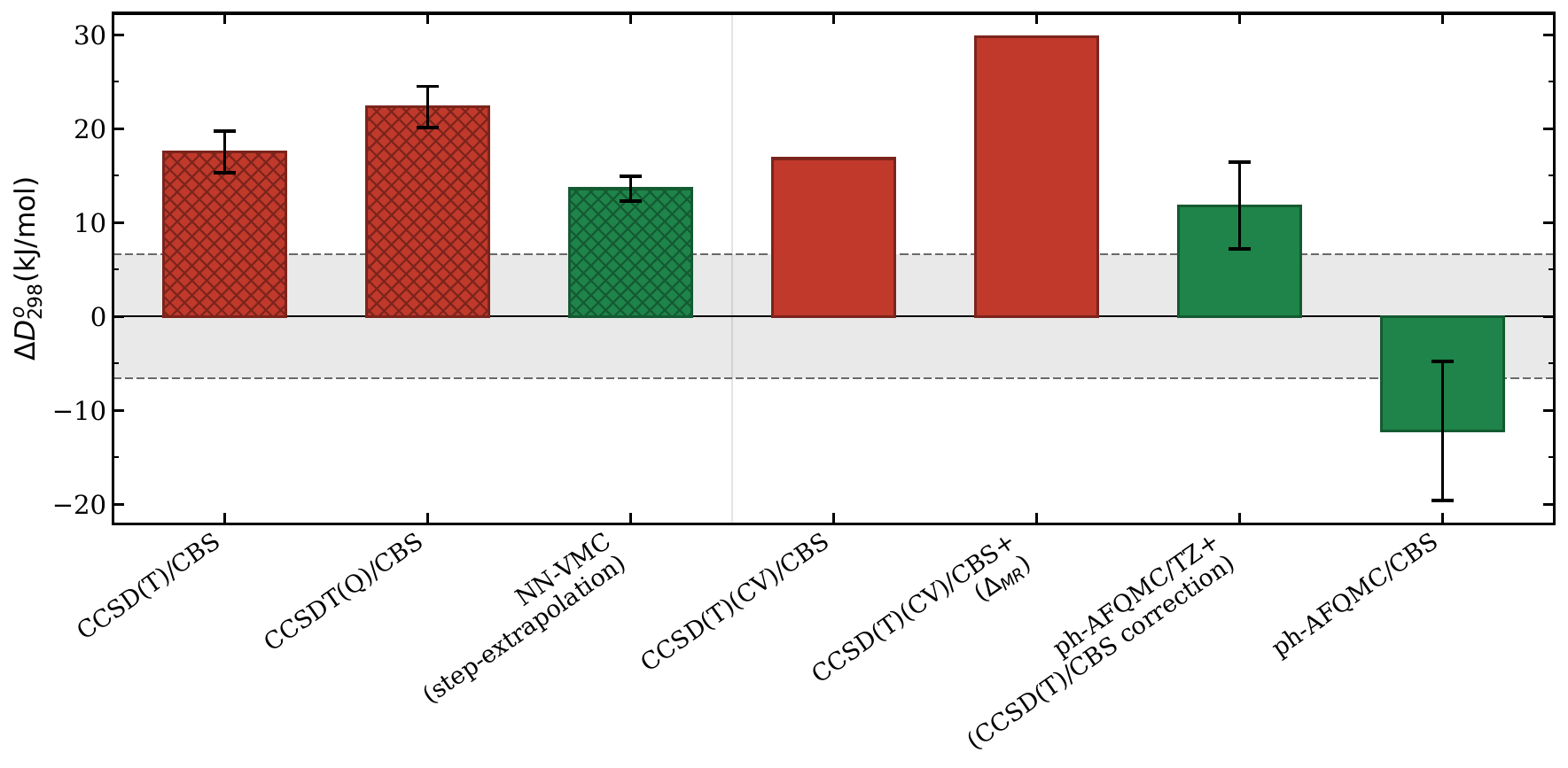}
    \caption{ Deviation of predicted NiH BDEs with respect to the average experimental value of 258.4(6.6) ~kJ/mol (See Table S3 in supporting information for list of experimental values). Our results are shown with cross-hatched bars and prior theory shown with solid colors.  Coupled cluster approaches are colored red, while QMC methods (NN-VMC and AFQMC) are marked green.
    The gray band around zero represents averaged experimental uncertainty.
    }
    \label{fig:dev-exp-ave-NiH}
\end{figure*}

We now examine hydride bond dissociation energies (BDE), $D_{298}^o$, obtained with various levels of theory, including coupled cluster and extrapolated neural network wavefunction data. 
Here we choose to focus the discussion on TiH and NiH.  Data for LiH and OH can be found in the section S7 of the supplementary information, where it can be seen that the results are essentially exact compared to prior experiment and other high-level theory and relevant explanations are provided in case of disagreement.

\paragraph{\textbf{TiH:}}

Among the $3d$ monohydrides, TiH provides an intermediate case, exhibiting moderate correlation effects while remaining computationally tractable. The experimental situation, however, remains unsettled. Two reported  values differ by approximately $50$~kJ/mol: \tihcrc  kJ/mol from the CRC Handbook \cite{CRC} and \textit{ca.}~159~kJ/mol from data reported in Lange's Handbook \cite{langes}. This discrepancy far exceeds the stated uncertainties and complicates direct experimental validation. Hence we compare our results with published high-accuracy theoretical predictions \cite{moltved-2019-ccsdt, aoto-2017-icMRCC, Shee-2019-AFQMC}.
These prior theoretical results for TiH bond dissociation energies, $D_{298}^o$, can be found in Table \ref{tab:energy_bind}, alongside experimental data and our own calculations.

In the present study, the ccECP-based CCSDT(Q)/CBS calculation yields a BDE of \BETiHccqcbs  kJ/mol, placing it higher than the stated CRC experimental value, but with very low statistical significance (1$\sigma$) due large uncertainty in the reported experimental estimate. A similar trend is observed in previous high-level coupled-cluster studies: Moltved \textit{et al.} reported a CCSD(T)/QZ value of $210.2$~kJ/mol, while Aoto \textit{et al.} obtained $215.7$~kJ/mol using icMRCC/CBS (with  MR corrections on CCSD(T)/CBS result).
Prior AFQMC found a somewhat lower value of $194.8(3.8)$~kJ/mol close to icMRCC without MR correction, which was reported at $200.2$ kJ/mol. 
Our own coupled cluster results largely comport with these prior studies, as we find dissociation energies of \BETiHcctcbs and \BETiHccqcbs~kJ/mol with CCSD(T)/CBS and CCSDT(Q)/CBS, respectively.
The NN-VMC results fall in the middle of the other reported theoretical values, where, infinite training step and zero variance extrapolations provide results within $\approx2 \sigma$ of each other, finding a BDE of \BETiHstep ~and ~\BETiHvar~kJ/mol. 
Since these results rely on extrapolation, it is worthwhile to consider the sensitivity to choices in the fitting process.  For this purpose, we have examined the consequences of the fitting form used in infinite step extrapolation, where the asymptotic energy is represented by an inverse power law (eq \ref{eq:ext-eq}).  As shown in supplementary Figure S11, selecting the decay exponent $\alpha$ over the range [$0.7-2.0$] shows robust performance over tested fitting windows, with the BDEs falling in a narrow range between $202.4(1.4)$ and $204.9(6.6)$~kJ/mol.
Altogether, the theoretical results provide a unified picture with nearly all methods predicting a bond dissociation energy of $\sim 195(4)$ or higher. 
This strongly suggests that the experimental results of \tihcrc (CRC\cite{CRC}) and 192.7(6.0) (Armentrout\cite{armentrout-1996-TiH-NiH}) are the most accurate.

\paragraph{\textbf{NiH:}}

NiH represents the most demanding system considered in this work due to its pronounced near-degeneracies and multireference character \cite{jiang-2012-niH-multi-ref}. Among the $3d$ hydrides examined by Moltved \textit{et al.} \cite{moltved-2019-ccsdt}, NiH was judged to have the largest absolute deviation between CCSD(T) predictions and experimental BDEs, 
marking it a particularly challenging case for electronic structure theories.
%making it a particularly stringent benchmark for neural-network variational Monte Carlo.
The experimental literature values summarized in Table \ref{tab:energy_bind} (more listed in supplementary Table S3) show a large scatter and uncertainties reaching several tens of kJ/mol, complicating direct assessment of theoretical accuracy. Tolbert and Beauchamp \cite{tolbert-1986-NiH} determined the NiH bond dissociation energy using ion-beam mass spectrometry. The heterolytic bond energy BDE(Ni$^+-$H$^-$) was obtained by bracketing hydride transfer reactions with reference donors of known thermochemistry, and the homolytic BDE was subsequently derived through a thermochemical cycle involving the ionization potential of Ni and the electron affinity of H yielding BDE(NiH) = 272(25) ~kJ/mol. Later, Armentrout and Beauchamp \cite{armentrout-1996-TiH-NiH} determined the NiH bond dissociation energy using guided ion beam tandem mass spectrometry. By measuring the onset energy of the reaction Ni$^+ +$ H$_2 \rightarrow$ NiH$^+ +$ H, they obtained the bond energy of NiH$^+$, which was subsequently converted to the neutral NiH BDE through a thermochemical cycle involving IP(Ni) and IP(NiH), giving 240(8.0)~kJ/mol, matching with the BDE listed in CRC Handbook \cite{CRC}, while an additional higher value of 289(13)~kJ/mol appears in Lange's Handbook \cite{langes}, though the primary experimental source underlying this compilation entry is not explicitly identified.

Due to the large spread in experimental results, we choose to compare theoretical BDEs with the average over experimental values.  Theoretical results referenced against the average of 258.4(6.6) ~kJ/mol are shown in Figure ~\ref{fig:dev-exp-ave-NiH}.
Within theoretical BDEs, we observe a consistent trend among high-level electronic structure methods. Both all-electron and ccECP-based coupled-cluster calculations, including single and multireference treatments, generally predict larger BDE values than the experimental central estimates. 
The CBS extrapolated CCSDT(Q) and CCSD(T) values (ours and those of  Aoto et al. \cite{aoto-2017-icMRCC}) fall between $\sim$275 and 280~kJ/mol. The inclusion of multireference corrections results in an even higher estimate of 290.7~kJ/mol \cite{aoto-2017-icMRCC}. 
A separate theoretical approach involving QMC methodology, is pursued by Shee \textit{et al.}, where ph-AFQMC, was applied to a set of 44 3d-TM diatomics. In most cases, the BDEs were computed from ph-AFQMC/TZ data with a residual CBS correction scaled via CCSD(T). For a select set of diatomics, including NiH, that showed large inconsistencies with experimental data, CBS extrapolation was additionally performed directly within ph-AFQMC (two point TZ-QZ). 
In the CCSD(T) based extrapolation case, (ph-AFQMC/CCSD(T)-CBS), the estimated BDE was 270.2(4.6)~kJ/mol, aligning closely with coupled cluster.
In contrast, the ph-AFQMC/CBS obtained solely with QMC data yielded a lower dissociation energy of 248.2(7.4)~kJ/mol.  
It is interesting to note that the methods that aim to more fully account for multireference effects differ by the most among prior theories, with a 42.5(7.4)~kJ/mol gap between the ph-AFQMC/CBS and CCSD(T)/CBS+$\Delta_{MR}$ results.

Compared to the just discussed theoretical methods, a potential advantage of the NN-VMC approach is that it both intrinsically includes multireference effects and naturally resides in the CBS (continuum) limit.
As mentioned \ref{subsec:TE_results}, we employed 160-determinant Psiformer ansatz on Ni and its hydride to investigate the bond dissociation question.  
At the final training point ($10^5$ steps), we find a BDE of \BENiHeotmoredet~kJ/mol.  At this point, the energy variance between Ni and NiH were nearly identical--a criterion that is often sought to improve the quality of energy differences.  
We find a similar result when applying infinite step extrapolation on the training data.  The extrapolated result of \BENiHstepmoredet~kJ/mol was further tested for robustness against parametric assumptions as in the case of TiH. The results from the sensitivity test are present in Figure ~\ref{fig:NiH-BDE-train-smooth-alpha} and discussed in detail in section \ref{subsubsec:step-extrap}, which align closely with CCSD(T)/CBS (ours and Aoto et al. \cite{aoto-2017-icMRCC} as well as with ph-AFQMC when combined with coupled cluster CBS corrections \cite{Shee-2019-AFQMC}. Out of the listed CBS extrapolated results in Table \ref{tab:energy_bind}, the icMR corrected CCSD(T) and ph-AFQMC/CBS estimates remain at the higher and lower end of the of theoretical estimates.

When considering the comparison between theory and experiment, there is no clear possibility to identify one most accurate theory to gauge against one most accurate experiment.  We can instead only assess consistency or inconsistency between theory and experiment in broader terms. If we assume that the experimental average makes for a reasonable basis of comparison, we find the deviation between the experimental average and individual theories with a joint uncertainty in the range of 1.0-5.3$\sigma$. Given the variability amongst experiments, and the differences observed between some high-level theories, it is clear that further work is warranted to fully resolve the challenge posed by NiH.

\section{Conclusions}
\label{sec:conclusions}

In this work, we assessed the accuracy and limitations of neural-network variational Monte Carlo (NN-VMC) for hydride bond dissociation energies across a sequence of systems ranging from simple main-group molecules (LiH and OH) to transition metal hydrides (TiH and NiH). In order to provide a meaningful assessment, we have taken particular care to both control for and accurately measure sources of systematic errors. 
And to achieve \textit{nearly} exact total energies via highly-accurate CCSDT(Q)/CBS, we probed and slightly improved the accuracy of some of the previous total energy calculations \cite{gani-2020-acc-engI}.

Partially due to their high degree of flexibility, neural-network wavefunctions continue to improve over very long training trajectories, sometimes leaving unrealized the full quality possible within a given ansatz.  For this reason, extrapolation approaches are often employed, in our study we include extrapolation to zero-variance and infinite-training length.  
While linearity is often assumed in these extrapolations, it is not guaranteed. 
In the case of zero-variance extrapolation, we have established a more formal basis to explain the approximate linearity empirically observed in QMC studies involving neural-network and other high quality wavefunctions. In addition, to control for systematic and statistical uncertainties in these procedures, we have developed robust regression techniques that further allow us to assess the sensitivity of the extrapolations. In all of our assessments, we employ both zero-variance and infinite training extrapolations, with significant agreement while still showing marginal, residual differences between the distinct approaches in more complicated systems with heavier elements.

The LiH and OH results provide useful reference cases. For these systems, we find that the end-of-training and extrapolated bond energies are essentially converged. The resulting BDE values are statistically consistent with our high quality CBS extrapolated \textit{ab initio} benchmarks such as FCI and CCSDT(Q) and also with published experimental BDEs in CRC Handbook \cite{CRC}. These systems therefore establish that for moderately correlated hydrides, the Psiformer ansatz in DeepQMC can reach the chemically relevant limit with little to no dependence on post-training extrapolation. Notably, the two extrapolation schemes give mutually consistent results for these simpler systems, validating the protocols before they are applied to harder cases.

The two transition metal hydrides studied here, TiH and NiH, have previously been identified as non-trivial cases when explored through various high-level electronic structure methods.
We find that, for Ti and TiH, the default 16-determinant Psiformer wavefunction reaches sub-milli-Hartree agreement with CCSD(T)/CBS by the end of training. Zero-variance and infinite step extrapolations lower the total energies by a further 1-3 mHa, 
at which point more than 99.5\% of the correlation energy has been recovered when compared with CCSDT(Q)/CBS as shown in Figure ~\ref{fig:cor_Ti_TiH}.
The two NN-VMC extrapolation schemes produce BDE ($D_{298}^o$), ~\BETiHvar ~and~\BETiHstep ~kJ/mol, respectively, which are $2\sigma$ apart. These values agree closely with the CRC estimate of 204.6(8.8) kJ/mol. While the BDE obtained from CCSDT(Q)/CBS, \BETiHccqcbs ~kJ/mol, remains close to the upper bound on the experimental estimate. 

Assuming our very fine scale for the correlation energy recovery, NiH appears to be significantly different from TiH and it comes as the most stringent benchmark considered here. With the default 16-determinant ansatz, NN-VMC recovers more correlation than finite-basis CCSD(T)/5Z, but the extrapolated total energies leave 1-2\% of the correlation energy unaccounted for. 
We show that substantially expanding the Psiformer ansatz to include 10$\times$ the number of determinants systematically improves the wavefunction to capture all but $\sim$0.5\% of the correlation energy even for this challenging highly multireference system (see Figure ~\ref{fig:cor_Ni_NiH} and supplementary Table S7).
We provide a detailed comparison of NiH bond dissociation between the still nascent NN-VMC approach and other high level theories previously applied to this challenging system.
The prior theories present an unusual spread among the estimated BDEs. While coupled cluster calculations with complete basis set extrapolation show relatively tight agreement between 275-280~kJ/mol, methodologies aiming to better account for multireference effects produce substantially higher and lower estimates. 
At these extremes, pure ph-AFQMC/CBS estimates a BDE of 248.2(7.4)~kJ/mol while CCSD(T)(CV)/CBS+$\Delta_{MR}$  predicts dissociation to occur at 290.7~kJ/mol. Careful application of the expanded Psiformer ansatz along with infinite-step extrapolation yields results closely in line with coupled cluster, finding a $D_{298}^o$(NiH) of \BENiHstepmoredet ~kJ/mol.

While existing experimental data are insufficient to discriminate among the theoretical predictions due to lack of agreement and high uncertainties, our results establish NN-VMC, in combination with correlation-consistent effective core potentials (ccECP), as quantitatively competitive with the best coupled-cluster and QMC methods for main-group and early transition-metal hydrides.
We also note that NN-VMC approaches provide samples of explicitly constructed variational wavefunctions and therefore this allows for direct variational calculations of arbitrary expectations without any further bias. However, this research angle is beyond the scope of the present study and it is left for future explorations.

\section*{Acknowledgments}

This research (A.S., L.M., J.T.K.) was supported by the U.S. Department of Energy, Office of Science, Basic Energy Sciences, Materials Sciences and Engineering Division, as part of the Computational Materials Sciences Program and Center for Predictive Simulation of Functional Materials. A portion (P. G. -- early conceptualization, discussion, writing) of this research was sponsored by the Laboratory Directed Research and Development Program of Oak Ridge National Laboratory, managed by UT-Battelle, LLC, for the US Department of Energy.  The authors thank Paul R. C. Kent for reading the manuscript and providing helpful suggestions.

This research used resources of the National Energy Research Scientific Computing Center (NERSC), a U.S. Department of Energy Office of Science User Facility operated under Contract No. DE-AC02-05CH11231.
An award of computer time was provided by the Innovative and Novel Computational Impact on Theory and Experiment (INCITE) program.
This research used resources of the Oak Ridge Leadership Computing Facility, which is a DOE Office of Science User Facility supported under Contract No. DE-AC05-00OR22725.

%%%~~~~~~~~~~~~~~~~~~~~~~~~~~~~~~~~~~~~~~~~~~~~~~~~~~~
\section*{Conflict of Interest}
The authors have no conflicts to disclose.
%%%~~~~~~~~~~~~~~~~~~~~~~~~~~~~~~~~~~~~~~~~~~~~~~~~~~~
\section*{Author Contributions}

\textbf{A. Shaikh:} NN-VMC and coupled cluster calculations (lead), technical development of extrapolation methods (equal), data analysis (equal), writing (equal). 
\textbf{L. Mitas:} data analysis (equal), writing (equal), mentorship (equal).
\textbf{P. Ganesh:} writing (equal), conceptualization (supporting).
\textbf{J.T. Krogel:} technical development of extrapolation methods (equal), data analysis (equal), writing (equal), mentorship (equal), conceptualization (lead).

%%%~~~~~~~~~~~~~~~~~~~~~~~~~~~~~~~~~~~~~~~~~~~~~~~~~~~
%\section*{DATA AVAILABILITY}
%The data supporting the findings of this study are available in the supplementary material 
%with full data hosted by the Materials Data Facility[link to be provided upon acceptance].
%%%~~~~~~~~~~~~~~~~~~~~~~~~~~~~~~~~~~~~~~~~~~~~~~~~~~~
\section*{REFERENCES}
\bibliographystyle{apsrev4-2}
\bibliography{main.bib}

@article{Carleo2017Science,
  author    = {Giuseppe Carleo and Matthias Troyer},
  title     = {Solving the Quantum Many-Body Problem with Artificial Neural Networks},
  journal   = {Science},
  volume    = {355},
  number    = {6325},
  pages     = {602--606},
  year      = {2017},
  month     = {feb},
  doi       = {10.1126/science.aag2302},
  url       = {https://doi.org/10.1126/science.aag2302},
  publisher = {American Association for the Advancement of Science}
}

@article{Pfau-2020-ferminet,
  title = {Ab initio solution of the many-electron Schr\"odinger equation with deep neural networks},
  author = {Pfau, David and Spencer, James S. and Matthews, Alexander G. D. G. and Foulkes, W. M. C.},
  journal = {Phys. Rev. Res.},
  volume = {2},
  issue = {3},
  pages = {033429},
  numpages = {20},
  year = {2020},
  month = {Sep},
  publisher = {American Physical Society},
  doi = {10.1103/PhysRevResearch.2.033429},
  url = {https://link.aps.org/doi/10.1103/PhysRevResearch.2.033429}
}

@article{Hermann-2020-paulinet,
  author    = {Jan Hermann and Zeno Sch{\"a}tzle and Frank No{\'e}},
  title     = {Deep-neural-network solution of the electronic Schr{\"o}dinger equation},
  journal   = {Nature Chemistry},
  volume    = {12},
  number    = {10},
  pages     = {891--897},
  year      = {2020},
  month     = oct,
  doi       = {10.1038/s41557-020-0544-y},
  url       = {https://doi.org/10.1038/s41557-020-0544-y},
  publisher = {Springer Nature}
}

@inproceedings{
vonGlehn-2022-psiformer,
title={A Self-Attention Ansatz for Ab-initio Quantum Chemistry},
author={Ingrid von Glehn and James S Spencer and David Pfau},
booktitle={The Eleventh International Conference on Learning Representations },
year={2023},
url={https://openreview.net/forum?id=xveTeHVlF7j}
}

@article{choo-2020-fermionicNQS,
  author  = {Kenny Choo and Antonio Mezzacapo and Giuseppe Carleo},
  title   = {Fermionic neural-network states for ab-initio electronic structure},
  journal = {Nature Communications},
  volume  = {11},
  pages   = {2368},
  year    = {2020},
  doi     = {10.1038/s41467-020-15724-9},
  url     = {https://doi.org/10.1038/s41467-020-15724-9},
  issn    = {2041-1723},
  publisher = {Springer Nature}
}

@article{scherbela-2022-weightsharing-mol,
  author    = {Michael Scherbela and Rafael Reisenhofer and Leon Gerard and Philipp Marquetand and Philipp Grohs},
  title     = {Solving the Electronic Schr{\"o}dinger Equation for Multiple Nuclear Geometries with Weight-Sharing Deep Neural Networks},
  journal   = {Nature Computational Science},
  volume    = {2},
  number    = {5},
  pages     = {331--341},
  year      = {2022},
  month     = may,
  doi       = {10.1038/s43588-022-00228-x},
  url       = {https://doi.org/10.1038/s43588-022-00228-x},
  issn      = {2662-8457}
}

@article{li-2022-deepsolid,
  title={Ab initio calculation of real solids via neural network ansatz},
  author={Li, Xiang and Li, Zhe and Chen, Ji},
  journal={Nature Communications},
  volume={13},
  number={1},
  pages={7895},
  year={2022},
  publisher={Nature Publishing Group UK London},
  url= {https://www.nature.com/articles/s41467-022-35627-1}
}

@inproceedings{
gerard-2022-goldstandard,
title={Gold-standard solutions to the Schr\"odinger equation using deep learning: How much physics do we need?},
author={Leon Gerard and Michael Scherbela and Philipp Marquetand and Philipp Grohs},
booktitle={Advances in Neural Information Processing Systems},
editor={Alice H. Oh and Alekh Agarwal and Danielle Belgrave and Kyunghyun Cho},
year={2022},
url={https://openreview.net/forum?id=nX-gReQ0OT}
}

@article{li-2022-ccECP-ML,
  author  = {Li, Xiang and Fan, Cunwei and Ren, Weiluo and Chen, Ji},
  title   = {Fermionic Neural Network with Effective Core Potential},
  journal = {Physical Review Research},
  volume  = {4},
  number  = {1},
  pages   = {013021},
  year    = {2022},
  doi     = {10.1103/PhysRevResearch.4.013021}
}

@article{pfau-2024-excited,
  title={Accurate computation of quantum excited states with neural networks},
  author={Pfau, David and Axelrod, Simon and Sutterud, Halvard and von Glehn, Ingrid and Spencer, James S},
  journal={Science},
  volume={385},
  number={6711},
  pages={eadn0137},
  year={2024},
  url={https://doi.org/10.1126/science.adn0137},
}

@article{scherbela-2024-transferable,
  author    = {Michael Scherbela and Leon Gerard and Philipp Grohs},
  title     = {Towards a Transferable Fermionic Neural Wavefunction for Molecules},
  journal   = {Nature Communications},
  volume    = {15},
  pages     = {120},
  year      = {2024},
  month     = jan,
  doi       = {10.1038/s41467-023-44216-9},
  url       = {https://doi.org/10.1038/s41467-023-44216-9},
  publisher = {Springer Nature}
}

@article{szabo-2024-excitedstatevmc,
  author    = {P. B. Szab{\'o} and Zeno Sch{\"a}tzle and Matthew Entwistle and Frank No{\'e}},
  title     = {An Improved Penalty-Based Excited-State Variational Monte Carlo Approach with Deep-Learning Ansatzes},
  journal   = {Journal of Chemical Theory and Computation},
  volume    = {20},
  number    = {18},
  pages     = {8128--8140},
  year      = {2024},
  month     = sep,
  doi       = {10.1021/acs.jctc.4c00678},
  url       = {https://doi.org/10.1021/acs.jctc.4c00678},
  publisher = {American Chemical Society}
}

@article{scherbela-2025-deeperwin,
  author    = {Michael Scherbela and Leon Gerard and Halvard Sutterud and W. M. C. Foulkes and Philipp Grohs},
  title     = {Transferable Neural Wavefunctions for Solids},
  journal   = {Nature Computational Science},
  volume    = {5},
  number    = {12},
  pages      = {1147--1157},
  year      = {2025},
  month     = dec,
  doi       = {10.1038/s43588-025-00872-z},
  url       = {https://doi.org/10.1038/s43588-025-00872-z},
  issn      = {2662-8457}
}

@article{cassella-2023-discovering,
  title={Discovering quantum phase transitions with fermionic neural networks},
  author={Cassella, Gino and Sutterud, Halvard and Azadi, Sam and Drummond, ND and Pfau, David and Spencer, James S and Foulkes, W Matthew C},
  journal={Physical review letters},
  volume={130},
  number={3},
  pages={036401},
  year={2023},
  publisher={APS},
  url={https://doi.org/10.1103/PhysRevLett.130.036401}
}

@article{ren-jichen-2023-NN-vmc-dmc,
  author    = {Weiluo Ren and Weizhong Fu and Xiaojie Wu and Ji Chen},
  title     = {Towards the Ground State of Molecules via Diffusion Monte Carlo on Neural Networks},
  journal   = {Nature Communications},
  volume    = {14},
  pages     = {1860},
  year      = {2023},
  month     = apr,
  doi       = {10.1038/s41467-023-37609-3},
  url       = {https://doi.org/10.1038/s41467-023-37609-3},
  publisher = {Springer Nature}
}

@article{kessler-2021-ANNQMC,
  author  = {Kessler, Johannes and Calcavecchia, Francesco and K{\"u}hne, Thomas D.},
  title   = {Artificial Neural Networks as Trial Wave Functions for Quantum Monte Carlo},
  journal = {Advanced Theory and Simulations},
  volume  = {4},
  pages   = {2000269},
  year    = {2021},
  doi     = {10.1002/adts.202000269}
}

@article{ceperley-1977-fermionvmc,
  author    = {David M. Ceperley and Geoffrey V. Chester and Malvin H. Kalos},
  title     = {Monte Carlo Simulation of a Many-Fermion System},
  journal   = {Physical Review B},
  volume    = {16},
  number    = {7},
  pages     = {3081--3099},
  year      = {1977},
  month     = oct,
  doi       = {10.1103/PhysRevB.16.3081},
  url       = {https://doi.org/10.1103/PhysRevB.16.3081},
  publisher = {American Physical Society}
}

@article{ceperley-1980-electrongas,
  author    = {David M. Ceperley and B. J. Alder},
  title     = {Ground State of the Electron Gas by a Stochastic Method},
  journal   = {Physical Review Letters},
  volume    = {45},
  number    = {7},
  pages     = {566--569},
  year      = {1980},
  month     = aug,
  doi       = {10.1103/PhysRevLett.45.566},
  url       = {https://doi.org/10.1103/PhysRevLett.45.566},
  publisher = {American Physical Society}
}

@article{umrigar-1988-optimized,
  author    = {C. J. Umrigar and Kenneth G. Wilson and John W. Wilkins},
  title     = {Optimized Trial Wave Functions for Quantum Monte Carlo Calculations},
  journal   = {Physical Review Letters},
  volume    = {60},
  number    = {17},
  pages     = {1719--1722},
  year      = {1988},
  month     = apr,
  doi       = {10.1103/PhysRevLett.60.1719},
  url       = {https://doi.org/10.1103/PhysRevLett.60.1719},
  publisher = {American Physical Society}
}

@article{foulkes-2001-qmcreview,
  author    = {W. M. C. Foulkes and L. Mitas and R. J. Needs and G. Rajagopal},
  title     = {Quantum Monte Carlo Simulations of Solids},
  journal   = {Reviews of Modern Physics},
  volume    = {73},
  number    = {1},
  pages     = {33--83},
  year      = {2001},
  month     = jan,
  doi       = {10.1103/RevModPhys.73.33},
  url       = {https://doi.org/10.1103/RevModPhys.73.33},
  publisher = {American Physical Society}
}

@article{needs-2020-CASINO,
  author    = {R. J. Needs and M. D. Towler and N. D. Drummond and P. L{\'o}pez R{\'i}os and J. R. Trail},
  title     = {Variational and Diffusion Quantum Monte Carlo Calculations with the CASINO Code},
  journal   = {The Journal of Chemical Physics},
  volume    = {152},
  number    = {15},
  pages     = {154106},
  year      = {2020},
  month     = apr,
  doi       = {10.1063/1.5144288},
  url       = {https://doi.org/10.1063/1.5144288},
  publisher = {AIP Publishing}
}

@article{aoto-2017-icMRCC,
  author  = {Aoto, Yasmin A. and de Lima Batista, Andr{\'e} P. and K{\"o}hn, Andreas and de Oliveira-Filho, Agnaldo G. S.},
  title   = {How To Arrive at Accurate Benchmark Values for Transition Metal Compounds: Computation or Experiment?},
  journal = {Journal of Chemical Theory and Computation},
  volume  = {13},
  number  = {11},
  pages   = {5291--5316},
  year    = {2017},
  doi     = {10.1021/acs.jctc.7b00688},
  url     = {https://doi.org/10.1021/acs.jctc.7b00688}
}

@article{Shee-2019-AFQMC,
  author  = {Shee, James and Rudshteyn, Benjamin and Arthur, Evan J. and Zhang, Shiwei and Reichman, David R. and Friesner, Richard A.},
  title   = {On Achieving High Accuracy in Quantum Chemical Calculations of 3d Transition Metal Systems: A Comparison of Auxiliary-Field Quantum Monte Carlo with Coupled Cluster, Density Functional Theory, and Experiment for Diatomic Molecules},
  journal = {Journal of Chemical Theory and Computation},
  volume  = {15},
  number  = {4},
  pages   = {2346–2358},
  year    = {2019},
  doi     = {10.1021/acs.jctc.9b00083},
  url     = {https://doi.org/10.1021/acs.jctc.9b00083}
}

@article{moltved-2019-ccsdt,
  author    = {Klaus A. Moltved and Kasper P. Kepp},
  title     = {The Metal Hydride Problem of Computational Chemistry: Origins and Consequences},
  journal   = {The Journal of Physical Chemistry A},
  volume    = {123},
  number    = {13},
  pages     = {2888--2900},
  year      = {2019},
  month     = apr,
  doi       = {10.1021/acs.jpca.9b02367},
  url       = {https://doi.org/10.1021/acs.jpca.9b02367},
  publisher = {American Chemical Society}
}

@article{fu-ji-chen-2024-var-ext,
  author    = {Weizhong Fu and Weiluo Ren and Ji Chen},
  title     = {Variance extrapolation method for neural-network variational Monte Carlo},
  journal   = {Machine Learning: Science and Technology},
  year      = {2024},
  volume    = {5},
  number    = {1},
  pages     = {015016},
  doi       = {10.1088/2632-2153/ad1f75},
  publisher = {IOP Publishing}
}

@article{scherbela-2025-inv-step,
  author       = {Michael Scherbela and Nicholas Gao and Philipp Grohs and Stephan G{\"u}nnemann},
  title        = {Accurate Ab-initio Neural-network Solutions to Large-Scale Electronic Structure Problems},
  journal      = {arXiv preprint},
  volume       = {},
  year         = {2025},
  eprint       = {2504.06087},
  archivePrefix= {arXiv},
  url          = {https://arxiv.org/abs/2504.06087}
}

@Misc{website,
  Title                    = {Pseudopotential Library},
  Note                     = {A community website for pseudopotentials/effective core potentials developed for high accuracy correlated many-body methods such as quantum Monte Carlo and quantum chemistry (last accessed Aug. 03, 2026). },

  Url                      = {https://pseudopotentiallibrary.org/},
}

@article{cizek-1966-cc,
  author    = {Ji{\v{r}}{\'\i} {\v{C}}{\'\i}{\v{z}}ek},
  title     = {On the Correlation Problem in Atomic and Molecular Systems. Calculation of Wavefunction Components in Ursell-Type Expansion Using Quantum-Field Theoretical Methods},
  journal   = {The Journal of Chemical Physics},
  volume    = {45},
  number    = {11},
  pages     = {4256--4266},
  year      = {1966},
  month     = dec,
  doi       = {10.1063/1.1727484},
  url       = {https://doi.org/10.1063/1.1727484},
  publisher = {AIP Publishing}
}

@article{kato-1957,
  author  = {Kato, Tosio},
  title   = {On the Eigenfunctions of Many-Particle Systems in Quantum Mechanics},
  journal = {Communications on Pure and Applied Mathematics},
  volume  = {10},
  number  = {2},
  pages   = {151--177},
  year    = {1957},
  doi     = {10.1002/cpa.3160100201}
}

@article{jastrow-1955,
  author  = {Jastrow, Robert},
  title   = {Many-Body Problem with Strong Forces},
  journal = {Physical Review},
  volume  = {98},
  number  = {5},
  pages   = {1479--1484},
  year    = {1955},
  doi     = {10.1103/PhysRev.98.1479}
}

@article{irikura-2007-ZPE,
  author    = {Karl K. Irikura},
  title     = {Experimental Vibrational Zero-Point Energies: Diatomic Molecules},
  journal   = {Journal of Physical and Chemical Reference Data},
  year      = {2007},
  volume    = {36},
  number    = {2},
  pages     = {389--397},
  doi       = {10.1063/1.2436891},
  publisher = {AIP Publishing}
}

@book{CRC,
  editor    = {Leah R McEwen},
  title     = {CRC Handbook of Chemistry and Physics},
  edition   = {107},
  publisher = {CRC Press},
  address   = {Boca Raton, FL},
  year      = {2026},
  url       = {https://www.routledge.com/CRC-Handbook-of-Chemistry-and-Physics/McEwen/p/book/9781041194620},
}

@article{ruscic-2002,
  author  = {Ruscic, Branko and Wagner, Albert F. and Harding, Lawrence B. and Asher, Robert L. and Feller, David and Dixon, David A. and Peterson, Kirk A. and Song, Yang and Qian, Xueming and Ng, C. Y. and Liu, Jinbo and Chen, Weiwen},
  title   = {On the Enthalpy of Formation of Hydroxyl Radical and Gas-Phase Bond Dissociation Energies of Water and Hydroxyl},
  journal = {The Journal of Physical Chemistry A},
  volume  = {106},
  number  = {11},
  pages   = {2727--2747},
  year    = {2002},
  doi     = {10.1021/jp013909s}
}

@misc{NIST,
  author       = {Alexander Kramida and Yuri Ralchenko and Joseph Reader and NIST ASD Team},
  title        = {NIST Atomic Spectra Database (ver. 5.12)},
  year         = {2024},
  howpublished = {National Institute of Standards and Technology, Gaithersburg, MD},
  note         = {Available online: https://physics.nist.gov/asd (last accessed Aug 03, 2026)},
  doi          = {10.18434/T4W30F}
}

@article{1-ccECP,
    author = {Bennett, M. Chandler and Melton, Cody A. and Annaberdiyev, Abdulgani and Wang, Guangming and Shulenburger, Luke and Mitas, Lubos},
    title = {A new generation of effective core potentials for correlated calculations},
    journal = {The Journal of Chemical Physics},
    volume = {147},
    number = {22},
    pages = {224106},
    year = {2017},
    month = {12},
    issn = {0021-9606},
    doi = {10.1063/1.4995643},
    url = {https://doi.org/10.1063/1.4995643},
}

@article{3-ccECP,
    author = {Annaberdiyev, Abdulgani and Wang, Guangming and Melton, Cody A. and Bennett, M. Chandler and Shulenburger, Luke and Mitas, Lubos},
    title = {A new generation of effective core potentials from correlated calculations: 3d transition metal series},
    journal = {The Journal of Chemical Physics},
    volume = {149},
    number = {13},
    pages = {134108},
    year = {2018},
    month = {10},
    issn = {0021-9606},
    doi = {10.1063/1.5040472},
    url = {https://doi.org/10.1063/1.5040472},
}

@article{4-ccECP,
    author = {Wang, Guangming and Annaberdiyev, Abdulgani and Melton, Cody A. and Bennett, M. Chandler and Shulenburger, Luke and Mitas, Lubos},
    title = {A new generation of effective core potentials from correlated calculations: 4s and 4p main group elements and first row additions},
    journal = {The Journal of Chemical Physics},
    volume = {151},
    number = {14},
    pages = {144110},
    year = {2019},
    month = {10},
    issn = {0021-9606},
    doi = {10.1063/1.5121006},
    url = {https://doi.org/10.1063/1.5121006},
}

@book{langes,
  editor    = {John A. Dean},
  title     = {Lange's Handbook of Chemistry},
  edition   = {15th},
  publisher = {McGraw-Hill},
  address   = {New York},
  year      = {1999},
  isbn      = {9780070163843},
  url       = {https://books.google.com/books/about/Lange_s_Handbook_of_Chemistry.html?hl=es&id=C7QPAQAAMAAJ}
}

@book{huber-herzberg,
  author    = {K. P. Huber and G. Herzberg},
  title     = {Molecular Spectra and Molecular Structure. IV. Constants of Diatomic Molecules},
  publisher = {Springer},
  address   = {New York},
  year      = {1979},
  isbn      = {9781475709612},
  doi       = {10.1007/978-1-4757-0961-2}
}

@article{gani-2020-acc-engI,
    author = {Annaberdiyev, Abdulgani and Wang, Guangming and Melton, Cody A. and Bennett, M. Chandler and Shulenburger, Luke and Mitas, Lubos},
    title = {A new generation of effective core potentials from correlated calculations: 3d transition metal series},
    journal = {The Journal of Chemical Physics},
    volume = {149},
    number = {13},
    pages = {134108},
    year = {2018},
    month = {10},
    issn = {0021-9606},
    doi = {10.1063/1.5040472},
    url = {https://doi.org/10.1063/1.5040472},
}

@Misc{MOLPRO,
  Title                    = {MOLPRO, version 2019.1, a package of ab initio programs},

  Author                   = {
H. J. Werner and
P. J. Knowles and
G. Knizia and
F. R. Manby and
M. Schutz
},
  Year                     = {2019},
  Url                      = {http://www.molpro.net},
  note                     = {(last accessed Aug. 03, 2026)},
}

@Misc{Mrcc,
  Title                    = {Mrcc, a quantum chemical program suite},

  Author                   = {M. Kallay and P. R. Nagy and Z. Rolik and D. Mester and G. Samu and J. Csontos and J. Csoka and B. P. Szabo and L. Gyevi-Nagy and I. Ladjanszki and L. Szegedy and B. Ladoczki and K. Petrov and M. Farkas and P. D. Mezei and and B. Hegely},

  Url                      = {https://www.mrcc.hu/}
}

@Article{extrapolation,
  Title                    = {Accurate ab initio-based molecular potentials: from extrapolation methods to global modelling},
  Author                   = {A J C Varandas},
  Journal                  = {Phys. Scr.},
  Year                     = {2007},

  Month                    = {aug},
  Number                   = {3},
  Pages                    = {C28--C35},
  Volume                   = {76},
  Doi                      = {10.1088/0031-8949/76/3/n04},
  Publisher                = {{IOP} Publishing},
  Url                      = {https://doi.org/10.1088%2F0031-8949%2F76%2F3%2Fn04}
}

@article{aqsa-2025-acc-eng-II,
    author = {Shaikh, Aqsa and Madany, Omar and Kincaid, Benjamin and Mitas, Lubos},
    title = {Accurate Atomic Correlation and Total Energies for
Correlation-Consistent Effective Core Potentials II: Rb–Xe
Elements},
    journal = {Journal of Chemical Theory and Computation},
    volume = {22},
    number = {9},
    pages = {4716-4727},
    year = {2026},
    month = {04},
    issn = {1549-9618},
    doi = {10.1021/acs.jctc.5c02129},
    url = {https://doi.org/10.1021/acs.jctc.5c02129},
}

@article{deepqmc,
    author = {Schätzle, Z. and Szabó, P. B. and Mezera, M. and Hermann, J. and Noé, F.},
    title = "{DeepQMC: An open-source software suite for variational optimization of deep-learning molecular wave functions}",
    journal = {The Journal of Chemical Physics},
    volume = {159},
    number = {9},
    pages = {094108},
    year = {2023},
    month = {09},
    issn = {0021-9606},
    doi = {10.1063/5.0157512},
    url = {https://doi.org/10.1063/5.0157512},
}

@article{2018-jax,
  author = {James Bradbury and Roy Frostig and Peter Hawkins and Matthew James Johnson and Chris Leary and Dougal Maclaurin and George Necula and Adam Paszke and Jake VanderPlas and Skye Wanderman-Milne and Qiao Zhang},
  title = {JAX: composable transformations of Python+NumPy programs},
  journal = {GitHub repository},
  year = {2018},
  note = {\url{https://github.com/jax-ml/jax}}
}

@InProceedings{2015-kfac,
  title = 	 {Optimizing Neural Networks with Kronecker-factored Approximate Curvature},
  author = 	 {Martens, James and Grosse, Roger},
  booktitle = 	 {Proceedings of the 32nd International Conference on Machine Learning},
  pages = 	 {2408--2417},
  year = 	 {2015},
  editor = 	 {Bach, Francis and Blei, David},
  volume = 	 {37},
  series = 	 {Proceedings of Machine Learning Research},
  address = 	 {Lille, France},
  month = 	 {07--09 Jul},
  publisher =    {PMLR},
  url = 	 {https://proceedings.mlr.press/v37/martens15.html},
}

@article{ve-ratio-LiNiO-Jaron-2026,
    author = {Ghaffar, Abdul and Zhang, Shenli and Wan, Liwen F. and Saritas, Kayahan and Reboredo, Fernando A. and Krogel, Jaron T.},
    title = {Many-Body Benchmark
of Electronic Charge and Spin
Densities for Li1–xNiO2},
    journal = {Journal of Chemical Theory and Computation},
    volume = {22},
    number = {9},
    pages = {4346-4357},
    year = {2026},
    month = {04},
    issn = {1549-9618},
    doi = {10.1021/acs.jctc.5c02097},
    url = {https://doi.org/10.1021/acs.jctc.5c02097},

}

@Misc{qmcpack_manual,
  Title                    = {{QMCPACK} User's Guide and Developer's Manual},
  Note                     = {Section ``Judging wavefunction optimization,'' (last accessed Aug. 15, 2026 },

  Url                      = {https://qmcpack.readthedocs.io/en/develop/analyzing.html},
}

@article{mitas-1991,
    author = {Mitáš, Luboš and Shirley, Eric L. and Ceperley, David M.},
    title = {Nonlocal pseudopotentials and diffusion Monte Carlo},
    journal = {The Journal of Chemical Physics},
    volume = {95},
    number = {5},
    pages = {3467-3475},
    year = {1991},
    month = {09},
    issn = {0021-9606},
    doi = {10.1063/1.460849},
    url = {https://doi.org/10.1063/1.460849},
}

@article{ceperly-1993-2d,
  author  = {Kwon, Yongkyung and Ceperley, David M. and Martin, Richard M.},
  title   = {Effects of Three-Body and Backflow Correlations in the Two-Dimensional Electron Gas},
  journal = {Physical Review B},
  volume  = {48},
  number  = {16},
  pages   = {12037--12046},
  year    = {1993},
  doi     = {10.1103/PhysRevB.48.12037}
}

@InProceedings{mitas-1993,
author="Mit{\'a}{\v{s}}, L.",
editor="Landau, David P.
and Mon, K. K.
and Sch{\"u}ttler, Heinz-Bernd",
title="Pseudopotential Quantum Monte Carlo for Large-Z Atom Systems",
booktitle="Computer Simulation Studies in Condensed-Matter Physics V",
year="1993",
publisher="Springer Berlin Heidelberg",
address="Berlin, Heidelberg",
pages="94--105",
isbn="978-3-642-78083-7",
url = "https://doi.org/10.1007/978-3-642-78083-7_8"
}

@article{theil-1950,
  author  = {Theil, Henri},
  title   = {A Rank-Invariant Method of Linear and Polynomial Regression Analysis},
  journal = {Proceedings of the Koninklijke Nederlandse Akademie van Wetenschappen},
  volume  = {53},
  pages   = {386--392},
  year    = {1950},
  url     = {https://doi.org/10.1007/978-94-011-2546-8_20}
}

@article{sen-1968,
  author  = {Sen, Pranab K.},
  title   = {Estimates of the Regression Coefficient Based on Kendall's Tau},
  journal = {Journal of the American Statistical Association},
  volume  = {63},
  number  = {324},
  pages   = {1379--1389},
  year    = {1968},
  doi     = {10.1080/01621459.1968.10480934}
}

@article{ransac-1981,
  author  = {Fischler, Martin A. and Bolles, Robert C.},
  title   = {Random Sample Consensus: A Paradigm for Model Fitting with Applications to Image Analysis and Automated Cartography},
  journal = {Communications of the ACM},
  volume  = {24},
  number  = {6},
  pages   = {381--395},
  year    = {1981},
  doi     = {10.1145/358669.358692}
}

@Inbook{armentrout-1996-TiH-NiH,
author="Armentrout, P. B.
and Kickel, Bernice L.",
editor="Freiser, Ben S.",
title="Gas-phase thermochemistry of transition metal ligand systems: reassessment of values and periodic trends",
bookTitle="Organometallic Ion Chemistry",
year="1996",
publisher="Springer Netherlands",
address="Dordrecht",
pages="1--45",
isbn="978-94-009-0111-7",
doi="10.1007/978-94-009-0111-7_1",
url="https://doi.org/10.1007/978-94-009-0111-7_1"
}

@article{tolbert-1986-NiH,
    author = {Tolbert, M. A. and Beauchamp, J. L.},
    title = {Homolytic and heterolytic bond dissociation energies of the second row group 8, 9, and 10 diatomic transition-metal hydrides: correlation with electronic structure},
    journal = {The Journal of Physical Chemistry},
    volume = {90},
    number = {21},
    pages = {5015-5022},
    year = {2002},
    month = {05},
    issn = {0022-3654},
    doi = {10.1021/j100412a029},
    url = {https://doi.org/10.1021/j100412a029},
}

@article{jiang-2012-niH-multi-ref,
  author  = {Jiang, Wanyi and DeYonker, Nathan J. and Wilson, Angela K.},
  title   = {Multireference Character for 3d Transition-Metal-Containing Molecules},
  journal = {Journal of Chemical Theory and Computation},
  year    = {2012},
  volume  = {8},
  number  = {2},
  pages   = {460--468},
  doi     = {10.1021/ct2006852},
  url     = {https://doi.org/10.1021/ct2006852}
}

\end{document}